%% file: main.tex
\documentclass[aps,prl,twocolumn,superscriptaddress,showpacs,final,floatfix,longbibliography]{revtex4-2}

\usepackage{orcidlink} 

\usepackage{graphicx}
\usepackage{dcolumn}
\usepackage{bm}
\usepackage{hyperref}
\usepackage{gensymb}
\usepackage[utf8]{inputenc}
\usepackage{braket}
\usepackage{amsmath}
\usepackage{color}
\usepackage{soul}
\usepackage{xfrac}
\usepackage{amssymb}
\usepackage{comment}
\usepackage[normalem]{ulem}
\usepackage{float}
\newcommand\redsout{\bgroup\markoverwith{\textcolor{red}{\rule[0.5ex]{2pt}{0.4pt}}}\ULon}

\usepackage{physics}
\usepackage{dsfont}

\usepackage{bbold} 

\soulregister\cite{7} 
\soulregister\ref{7} 
\soulregister\eqref{7} 

\newcommand{\ignore}[1]{}

\newcommand{\be}{\begin{equation}}
\newcommand{\ee}{\end{equation}}
\newcommand{\eea}{\end{eqnarray}}
\newcommand{\bea}{\begin{eqnarray}}

\newcommand{\av}[1]{\ensuremath{\langle{#1} \rangle}}

\usepackage{cleveref}
\crefname{equation}{Eq.}{Eqs.}
\crefname{observation}{Obs.}{Obs.}
\crefname{corollary}{Corollary}{Corollaries}
\crefname{lemma}{Lemma}{Lemmata}
\crefname{proof}{Proof}{Proofs}
\creflabelformat{proof}{#2proof#3}
\crefname{remark}{Remark}{Remarks}
\crefname{prop}{Proposition}{Propositions}

\begin{document}


\title{Experimental certification of high-dimensional entanglement with randomized measurements}

\author{Ohad Lib$^\dagger$ \orcidlink{0000-0002-2010-0514}}
\email{ohad.lib@mail.huji.ac.il}
\affiliation{Racah Institute of Physics, The Hebrew University of Jerusalem, Jerusalem, 91904, Israel}

\author{Shuheng Liu \orcidlink{0000-0001-7130-1888}}
\thanks{O. Lib and S. Liu contributed equally to this work.}
\affiliation{State Key Laboratory for Mesoscopic Physics, School of Physics, Frontiers Science Center for Nano-optoelectronics, Peking University, Beijing 100871, China}
\affiliation{Vienna Center for Quantum Science and Technology, Atominstitut, TU Wien,  1020 Vienna, Austria}

\author{Ronen Shekel \orcidlink{0000-0003-1851-5785}}
\affiliation{Racah Institute of Physics, The Hebrew University of Jerusalem, Jerusalem, 91904, Israel}

\author{Qiongyi He \orcidlink{0000-0002-2408-4320}}
\email{qiongyihe@pku.edu.cn}
\affiliation{State Key Laboratory for Mesoscopic Physics, School of Physics, Frontiers Science Center for Nano-optoelectronics, Peking University, Beijing 100871, China}
\affiliation{Collaborative Innovation Center of Extreme Optics, Shanxi University, Taiyuan, Shanxi 030006, China}
\affiliation{Hefei National Laboratory, Hefei 230088, China}

\author{Marcus Huber \orcidlink{0000-0003-1985-4623}}
\affiliation{Vienna Center for Quantum Science and Technology, Atominstitut, TU Wien,  1020 Vienna, Austria}
\affiliation{Institute for Quantum Optics and Quantum Information (IQOQI), Austrian Academy of Sciences, 1090 Vienna, Austria}

\author{Yaron Bromberg \orcidlink{0000-0003-2565-7394}}
\affiliation{Racah Institute of Physics, The Hebrew University of Jerusalem, Jerusalem, 91904, Israel}

\author{Giuseppe Vitagliano \orcidlink{0000-0002-5563-3222}}
\email{giuseppe.vitagliano@tuwien.ac.at}
\affiliation{Vienna Center for Quantum Science and Technology, Atominstitut, TU Wien,  1020 Vienna, Austria}

\begin{abstract}
High-dimensional entangled states offer higher information capacity and stronger resilience to noise compared with two-dimensional systems. However, the large number of modes and sensitivity to random rotations complicate experimental entanglement certification. Here, we experimentally certify three-dimensional entanglement in a five-dimensional two-photon state using 800 Haar-random measurements implemented via a 10-plane programmable light converter. We further demonstrate the robustness of this approach against random rotations, certifying high-dimensional entanglement despite arbitrary phase randomization of the optical modes. This method, which requires no common reference frame between parties, opens the door for high-dimensional entanglement distribution through long-range random links. \end{abstract}

\maketitle

{\it Introduction.---}From non-locality \cite{freedman1972experimental,aspect1982experimental} to quantum computation \cite{jozsa2003role} and communication \cite{ekert1991quantum}, entanglement stands at the basis of quantum mechanics and technology. Therefore, the certification and quantification of entanglement have been the subject of intensive research over the past decades, aiming to achieve theoretically sound and experimentally efficient entanglement certification schemes \cite{guhne2009entanglement,friis2019entanglement}.

Recently, the higher information capacity and stronger resilience to noise \cite{erhard2020advances} of high-dimensional, qudit, entangled states have triggered considerable research on the certification and quantification of entanglement in high dimensions~\cite{SchmidtBarbaraPRA2000,ErkerQuantum2017quantifyinghigh,HuberEntropyPRA2013,HuberStructurePRL2013,Liu2024bounding,ResourceMorelliPRL2023,EckerOvercomingPRX2019,HuPathwaysPRL2021,KrennGenerationNAS2014,BavarescoMeasurementsNP2018,SchneelochQuantifyingNC2019,HerreraHighQuantum2020,HuberHighPRL2018,KlocklCharacterizingPRA2015}.majority of these methods require control and knowledge over the experimentally implemented measurement bases, which are often unavailable when manipulating and distributing high-dimensional entanglement encoded in multiple optical modes.

To relax the need to precisely measure each qudit in specific bases, entanglement certification methods based on randomized measurements have been proposed via negativity of the partial transpose~\cite{Zhou2020,Elben_2020b}, or directly with moments of randomized correlations~\cite{Tran_2015,Tran_2016,Dimi__2018,Saggio_2019,Ketterer_2019,Ketterer_2020,Knips_2020,BoundImaiPRL2021,Ketterer_2022}. More recently, such methods have been extended to even certify the entanglement-dimensionality~\cite{liu2023characterizing,ProbingWyderkaPRXQ2023}. However, while such methods have been demonstrated experimentally for two-dimensional and bound entangled states~\cite{wyderka2023complete,zhang2023experimental}, the difficulty of realizing Haar-randomized measurements in high dimensions hindered their realization for experimentally certifying high-dimensional entanglement.

In this letter, we experimentally certify three-dimensional entanglement using randomized measurements. To this end, we generate a five-dimensional state encoded in the spatial modes of two entangled photons and perform 800 Haar-random measurements. We implement the randomized measurements using a $10$-plane light converter based on optimized phase masks separated by free-space propagation, which are programmed to realize Haar-random unitary transformations on the spatial modes of both photons. From these measurements, we obtain 
randomized correlations and estimate second- and fourth-order moments to certify genuine three-dimensional entanglement using the methods of Ref.~\cite{liu2023characterizing}. By adding random relative phases to the modes encoding the state, we further demonstrate the robustness of this entanglement certification method against unknown random local rotations of the quantum state, opening the door for the certification of high-dimensional entanglement distributed through random quantum links.

\begin{figure}[ht!]
\centering
\includegraphics[width=0.97\columnwidth]{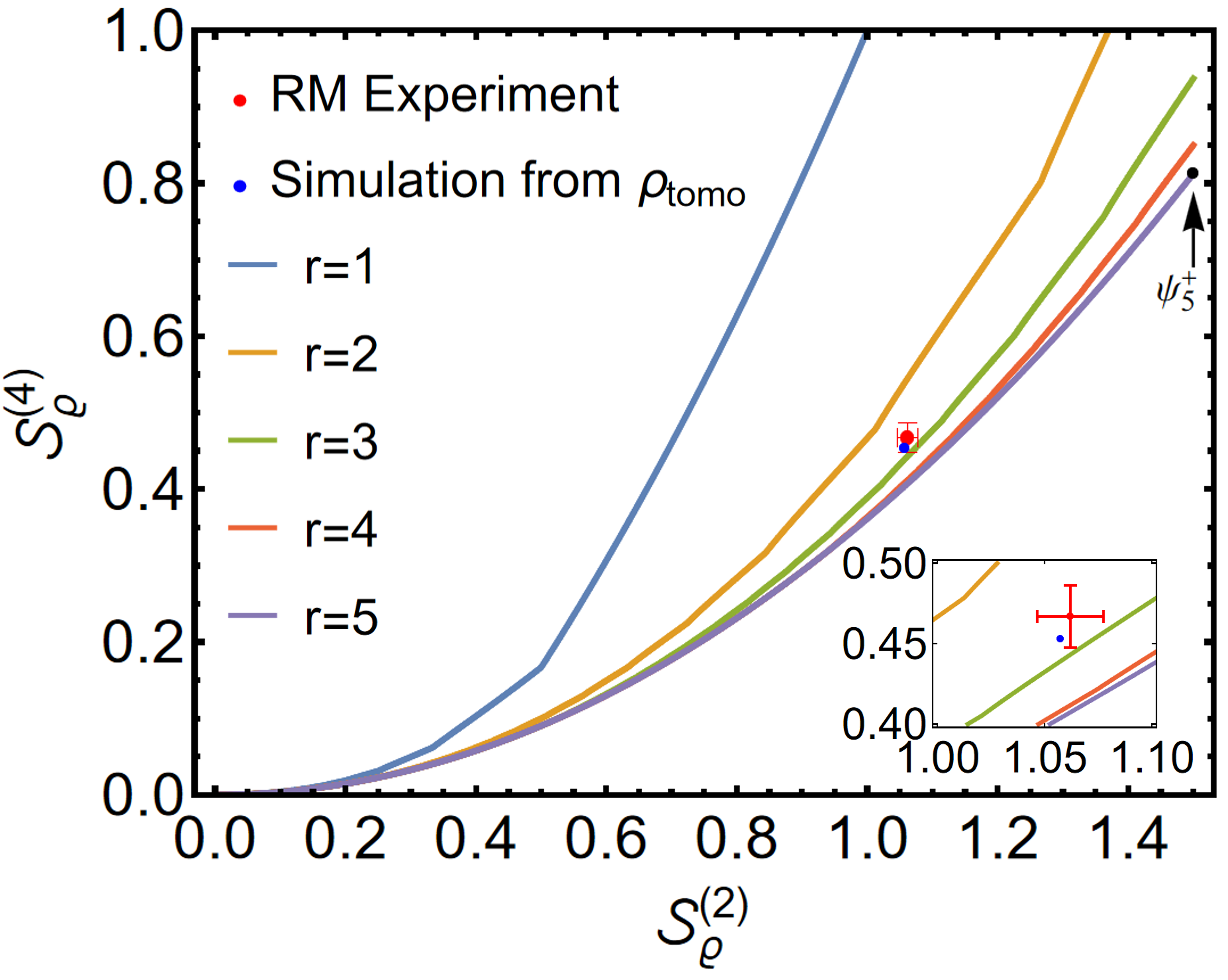}
\caption{Entanglement certification with randomized measurements. The solid lines in different colors represent lower bounds corresponding to different Schmidt numbers $r$. A point below the $r$-th curve certifies $(r+1)$-dimensional entanglement. The black point represents the ideal maximally entangled state on the $(\bar{\mathcal{S}}_{\varrho}^{(2)} , \bar{\mathcal{S}}_{\varrho}^{(4)})$ plane. The red point, with error bars representing 2 standard deviations, shows experimental data from randomized measurements. The blue point represents a simulation using the reconstructed density matrix $\varrho_{\text{tomo}}$ from tomography and the same $800$ chosen unitaries. 
Inset: Zoom around the data point.}
\label{fig:2}
\end{figure}

{\it Theoretical Methods.---}The certification of high-dimensional entanglement using randomized measurements relies on a criterion based on the cross-correlation matrix. Consider a bipartite state with equal dimensions $d$ and the $su(d)$ basis $\boldsymbol{g}=\{g_k\}_{k=1}^{d^2-1}$ with normalization $\operatorname{Tr}(g_k g_l)=\delta_{kl}$. The cross-correlation matrix has elements given by $\left(\mathfrak{X}_{\varrho}\right)_{k l}=\operatorname{Tr}(g_k\otimes g_l \varrho)$. For any state with Schmidt number $r$, the trace norm of $\mathfrak{X}_{\varrho}$, i.e. the sum of its singular values $\epsilon_k$ is upper-bounded by~\cite{liu2023characterizing}
\begin{equation}\label{eq:TraceNormCriterion}
\operatorname{tr}\left|\mathfrak{X}_{\varrho}\right|:=\sum_{k=1}^{d^2-1} \epsilon_k \leq r-\frac{1}{d} ,
\end{equation}
where the trace norm is defined as $\tr|A|=\tr\sqrt{A^\dagger A}$.
This inequality is invariant under local changes of reference frame and can be used to find nontrivial constraints on moments of $su(d)$ Bloch sphere integrals:
\begin{equation}
    \mathcal{S}_{\varrho}^{(m)}:=N \int d \hat{\boldsymbol{\alpha}} \int d \hat{\boldsymbol{\beta}}\left[\av{g_{\hat{\boldsymbol{\alpha}}} \otimes g_{\hat{\boldsymbol{\beta}}}}_{\varrho}\right]^m ,
\end{equation}
where the correlations $\av{g_{\hat{\boldsymbol{\alpha}}} \otimes g_{\hat{\boldsymbol{\beta}}}}_{\varrho}=\hat{\boldsymbol{\alpha}} \mathfrak{X}_{\varrho} \hat{\boldsymbol{\beta}}^T$ are obtained by multiplying the cross-correlation matrix with the unit $(d^2-1)-$dimensional vectors $\hat{\boldsymbol{\alpha}}$ and $\hat{\boldsymbol{\beta}}$ defining two directions in the $su(d)$ Bloch sphere and the integral is performed over all such directions.
Employing the constraint given in \cref{eq:TraceNormCriterion}, the $(\mathcal{S}_{\varrho}^{(2)} , \mathcal{S}_{\varrho}^{(4)})$-plane is partitioned into different regions corresponding to different Schmidt numbers (cf.~\cref{fig:2}). Here, we also note that the curves in \cref{fig:2} are known analytically~\cite{liu2023characterizing,ProbingWyderkaPRXQ2023}. In turn, 
such moments can be estimated from their relation with the randomized correlators 
\begin{equation}
    \mathfrak{R}_{\varrho}^{(m)}=\int {\mathrm d} U_a {\mathrm d} U_b \av{\left(U_a \otimes U_b\right) M \otimes M\left(U_a \otimes U_b\right)^{\dagger}}^{m} ,
\end{equation}
where $M$ is a given observable and
$U_a$ and $U_b$ are random unitaries sampled from the Haar measure ${\mathrm d} U$, over which the integral is performed. 
We emphasize once more that these measurements avoid the need to align the reference frames of the two parties.
By finding the estimated point in the $(\mathcal{S}_{\varrho}^{(2)}, \mathcal{S}_{\varrho}^{(4)})$-plane corresponding to the experimentally prepared quantum state and
comparing it to the boundary curves corresponding to the different Schmidt numbers we can then certify the minimal entanglement dimensionality of our quantum state.
See Supplemental Material for further details~\cite{supp}.

{\it Experiment.---}In the experiment, we prepare a five-dimensional statevia spontaneous parametric down-conversion (SPDC) (see Supplementary information~\cite{supp}). The five-dimensional state of each photon is encoded using five spatial modes, 
defined by five pairs of apertures placed at the far-field of the SPDC generation plane. Thanks to transverse momentum conservation in the SPDC process, when the apertures are placed symmetrically around the optical axis, the photons that pass the apertures should be found in an entangled state 
\begin{equation}\label{eq:MESideal}
\ket{\psi^+_5}=\frac{1}{\sqrt{5}} (\ket{11'}+\ket{22'}+\ket{33'}+\ket{44'}+\ket{55'}) ,    
\end{equation}
where $|ii'\rangle$ denotes the pair of modes defined by apertures $i$ and $i'$ placed symmetrically around the optical axis. 

\begin{figure}[ht!]
 \centering
 \includegraphics[width=0.97\columnwidth]{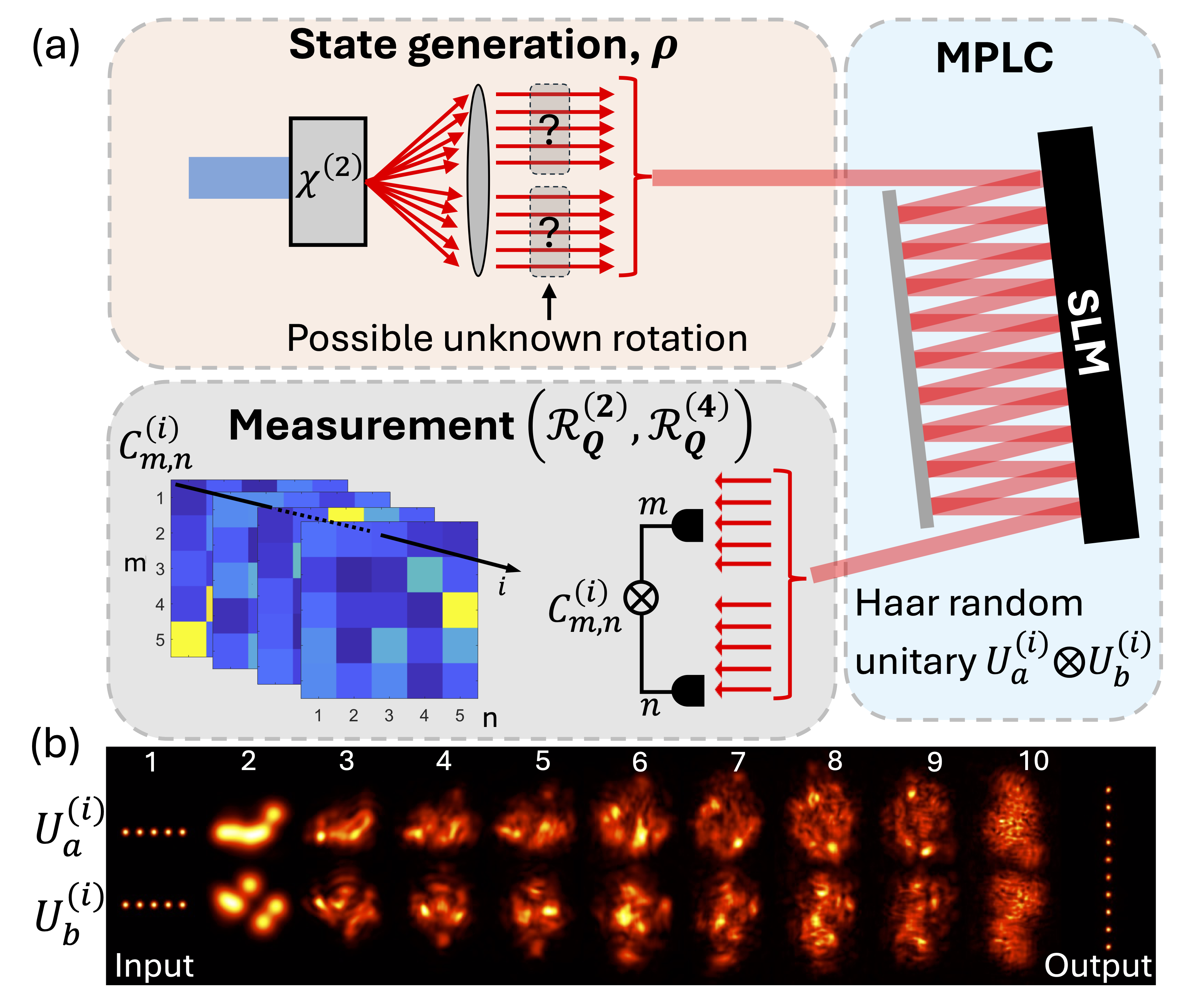}
 \caption{Scheme and experimental setup. (a) Photon pairs are generated via spontaneous parametric down-conversion (SPDC) by pumping a nonlinear crystal using a classical pump beam. We set the phase-matching conditions of the SPDC process to generate spatially entangled photons over five pairs of spots defined by ten apertures placed at the far field of the SPDC crystal. An arbitrary quantum link can potentially manifest as unknown rotations that act on the five-dimensional space of each photon. A 10-plane light converter consisting of a phase-only spatial light modulator (SLM) and mirror (gray) is programmed to apply different Haar-random unitary transformations $(U_a^{(i)} \otimes U^{(i)}_b)$ on the modes of each photon. Coincidence counting $C^{(i)}_{m,n}$ between all modes $m,n$ at the output of the light converter, measured for 800 transformations ($i= 1,\dots,800$), provides access to the distribution of randomized correlations in the state. Entanglement is then certified using the second- and fourth-order moments of this distribution (see text). (b) A simulation showing the total intensity of all modes at each plane of the light converter illustrates the five modes encoding the state of each photon and their random mixing in the light converter, for one of the transformations $(U_a^{(i)} \otimes U^{(i)}_b)$ used. The maximal intensity at each plane is normalized to one.}
\label{fig:1}
\end{figure}

The generated state can potentially undergo unknown random rotations before entanglement certification (cf. \cref{fig:1}a), preventing the use of standard entanglement witnesses~\cite{guhne2009entanglement,friis2019entanglement,lib2022quantum}. To certify the dimension of entanglement despite the potentially unknown rotations of the state we
apply the criterion described above, which requires the ability to perform five-dimensional Haar-random unitary transformations that mix the modes defined by the five spots of each photon (\cref{fig:1}b).

To this end, we use multi-plane light conversion (MPLC)\cite{morizur2010programmable}, where multiple phase masks separated via free-space propagation are used to perform arbitrary transformations on spatial modes of either classical\cite{labroille2014efficient,fontaine2019laguerre,lib2024high} or quantum\cite{brandt2020high,hiekkamaki2021high,lib2022processing,lib2024resource,lib2024high} light. We implement a 10-plane light converter built in a multi-pass configuration between a single spatial light modulator (SLM) and a mirror\cite{lib2024resource,lib2024high,lib2024building}, and use a wavefront matching algorithm\cite{fontaine2019laguerre} to design phase masks that realize Haar-random unitaries (\cref{fig:1}b). Following each random unitary 
$U^{(i)}_a \otimes U^{(i)}_b$, we perform coincidence measurements and obtain the correlations between the photons $C^{(i)}_{m,n}$ in all spatial modes $m,n$.

Using this set of randomized measurements, we turn to certify the dimension of entanglement in our state by estimating the second- and fourth-order moments of the Bloch sphere integrals through the relations 
\begin{equation}
\begin{aligned}
\mathcal{S}_{\varrho}^{(2)} &=(d+1)^2 \mathfrak{R}_{\varrho}^{(2)}, \\
\mathcal{S}_{\varrho}^{(4)} &=\frac{(d+1)^2\left(d^2+1\right)^2}{9(d-1)^2} \mathfrak{R}_{\varrho}^{(4)} ,
\end{aligned}
\end{equation}
which can be in turn estimated via a finite set of samples (in our case $N_{\mathrm{tot}}=800$) of the quantities $\av{\left(U^{(i)}_a \otimes U^{(i)}_b\right) M \otimes M\left(U^{(i)}_a \otimes U^{(i)}_b\right)^{\dagger}}$.
More details about this estimation are given in the Supplemental Material~\cite{supp}.
Our experimental result and the bounds for different entanglement dimensionalities are presented in \cref{fig:2}. We obtain the estimated values of $(\bar{\mathcal{S}}_{\varrho}^{(2)} \pm 2\sigma(\bar{\mathcal{S}}_{\varrho}^{(2)}), \bar{\mathcal{S}}_{\varrho}^{(4)} \pm 2\sigma(\bar{\mathcal{S}}_{\varrho}^{(4)}))=(1.06\pm 0.02,0.47\pm 0.02)$, where we take the mean value with an error estimated with two standard deviations. 
With these values, three-dimensional entanglement is certified in our five-dimensional state~\cite{liu2023characterizing}, demonstrating the ability to experimentally certify high-dimensional entanglement with randomized measurements.

To better understand why the experimental results differ from the expected moments for the ideal maximally entangled state \eqref{eq:MESideal}, we use the same experimental setup to perform quantum state tomography and reconstruct the density matrix of our state (See supplementary information~\cite{supp} \cref{figS:rhoTomoWeighted}). Using the tomographically-reconstructed density matrix $\varrho_{\text{tomo}}$ and the known $800$ unitary transformations, we estimate the expected moments $(\mathcal{S}_{\varrho}^{(2)} , \mathcal{S}_{\varrho}^{(4)})$ from a numerical analysis (\cref{fig:2}). Both the moments of the measured random correlations and the reconstructed density matrix can be reasonably explained using a dephasing model resulting from fluctuating phases between the different spatial modes on the level of $\pm 0.53$ radians. While we have not identified the exact cause of such phase fluctuations, they can potentially result from various imperfections in our experimental setup, such as wavelength instabilities in our classical pump beam, phase stability of the experimental setup, or imperfections in the MPLC. In addition to these imperfections, the main factor limiting the certified dimension of entanglement was the finite number and choice of measurement bases, which could be overcome by increasing the number of measurements (see Supplemental Material~\cite{supp} and discussion below).

Finally, we demonstrate the robustness of entanglement certification using randomized measurements for entangled photons distributed through random links. While arbitrary unitary rotations do not degrade the entanglement, they scramble the quantum correlations, making entanglement certification difficult~\cite{lib2022quantum}. This is particularly important for high-dimensional entanglement in the spatial domain, which is highly sensitive to scattering and aberrations in quantum links caused by atmospheric turbulence~\cite{krenn2015twisted}, cross-talk in multi-mode fibers~\cite{valencia2020unscrambling,shekel2023shaping}, or path-length differences in multi-core fibers~\cite{hu2020efficient}. While the effect of scattering and aberrations can potentially be mitigated via wavefront-shaping techniques that compensate for the action of the random link~\cite{wolterink2016programmable,defienne2016two,defienne2018adaptive,lib2020real,valencia2020unscrambling,shekel2024shaping}, their implementation incurs significant experimental overhead, requiring additional light sources, SLMs, and fast feedback mechanisms~\cite{lib2022quantum}.

\begin{figure}[ht!]
\centering
\includegraphics[width=0.97\columnwidth]{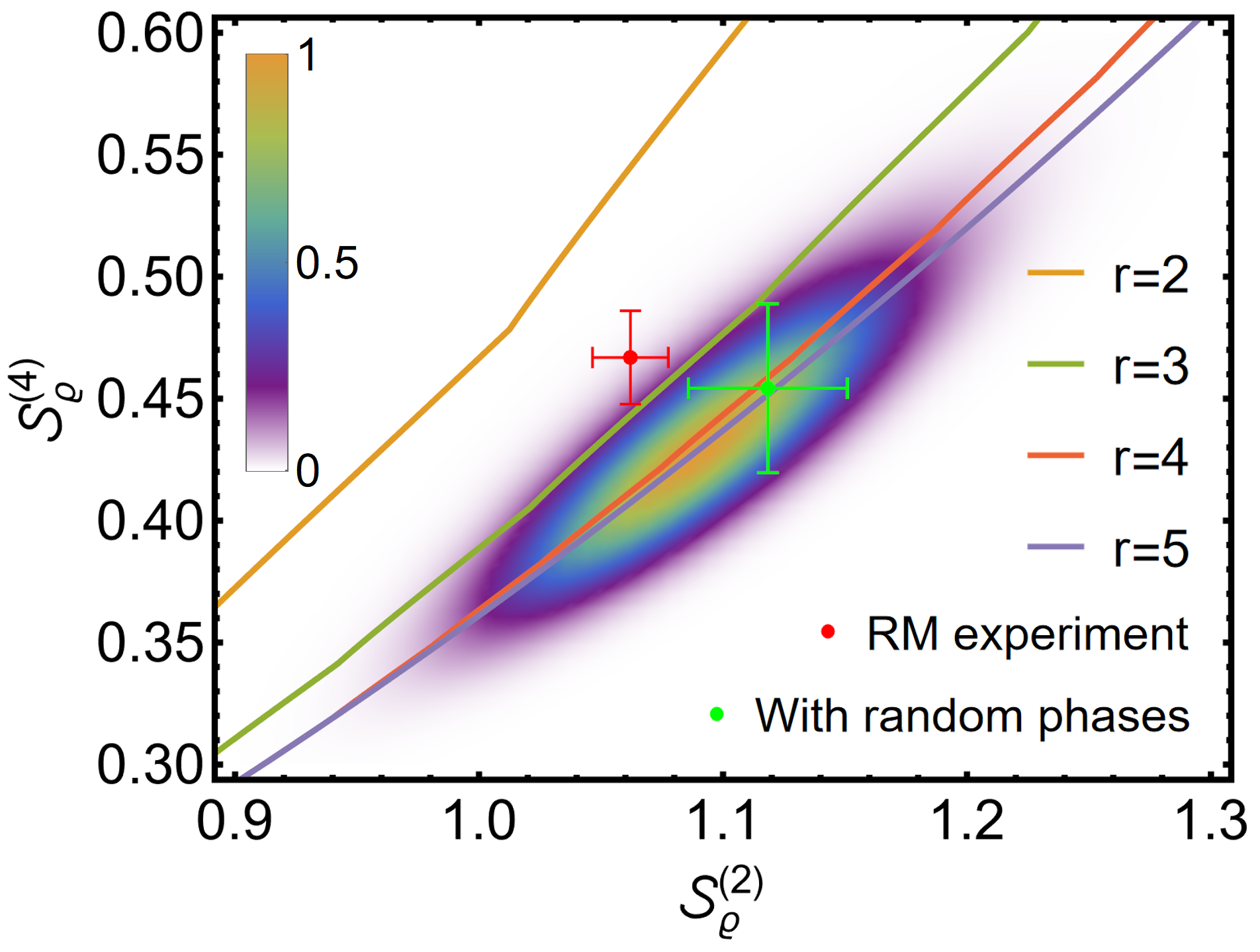}
\caption{Robustness to arbitrary state rotations. 
Experimental data points with (green) and without (red) random phases. Error bars represent 2 standard deviations. With the added random phases, it seems as if even the Schmidt number $r=4$ can be certified. The finite number of sampled unitaries explains the difference between the two results (green and red). The histogram of $10^6$ samples represents a simulation of results from random phases. The densities are normalized to the highest value (yellow = 1). White represents zero density. 
}
\label{fig:3}
\end{figure}

The advantage of entanglement certification via random measurements is precisely to be insensitive to such random rotations of the state during its distribution. To further demonstrate this advantage in practical situations, we use the first plane of the MPLC to emulate a simple random link where a different random phase is applied to each input spot and for each unitary, mimicking, for example, the case of entanglement distribution through multi-core fibers~\cite{hu2020efficient} (\cref{fig:1}). While these random phases scramble the correlations between the photons in specific bases (\cref{figS:DFTData}), high-dimensional entanglement is still certified using randomized measurements (\cref{fig:3}). One should note, however, that there is a difference in the measured $(\mathcal{S}_{\varrho}^{(2)} , \mathcal{S}_{\varrho}^{(4)})$ values between the measurements with and without the random phases. As we show in \cref{fig:3} and in the Supplemental Material~\cite{supp}, this difference results from the finite number of sample unitaries (which is $N_{\mathrm{tot}}=800$) and is compatible with our estimated density matrix. The histogram of possible $(\mathcal{S}_{\varrho}^{(2)} , \mathcal{S}_{\varrho}^{(4)})$ values for our reconstructed density matrix and different sets of random phases is presented in \cref{fig:3}. This finite sampling of unitaries could also potentially explain why three-dimensional entanglement rather than five-dimensional entanglement is certified (see Supplemental Material~\cite{supp}).

{\it Conclusions.---}We have presented the first experimental demonstration of high-dimensional entanglement certification with randomized measurements. We have further demonstrated the robustness of this method against the scrambling of correlations due to random accumulated phases of the different spatial modes, which is an important advantage of our method for potential applications in detecting entanglement between spatially distant parties.
Currently, the main limitation of the experimental implementation of this method is the large number of required randomized measurements, but we expect future research to reduce this experimental overhead. For example, one can further investigate the direct use of random links for randomized measurements without additional hardware. 
Beyond these prospects, we believe our current demonstration already opens the door to direct entanglement certification of high-dimensional entanglement through long-range random links without establishing a common reference frame.

\begin{acknowledgments}
{\it Acknowledgments.---} O.L. acknowledges the support of the Clore Scholars Programme of the Clore Israel Foundation. S.L. acknowledges the China Postdoctoral Science Foundation (No. 2023M740119). R.S acknowledges the support of the Israeli Council for Higher Education, and of the HUJI center for nanoscience and nanotechnology. This research was supported by the Zuckerman STEM Leadership Program, the Israel Science Foundation (grant No. 2497/21), the National Natural Science Foundation of China (Grants No. 12125402, No. 12405005). G.V. acknowledges financial support from the Austrian Science Fund (FWF) through the grants P 35810-N and P 36633-N (Stand-Alone). M.H. acknowledges funding from the European Research Council (Consolidator grant ‘Cocoquest’ 101043705) and the Horizon-Europe research and innovation programme under grant agreement No 101070168 (HyperSpace).

\end{acknowledgments}

\bibliographystyle{apsrev4-2}
\bibliography{refs}

\appendix
\begin{widetext}
\newpage
\include{SI.tex}
\end{widetext}

\end{document}

%% file: SI.tex
\def\thefigure{S\arabic{figure}}
\setcounter{figure}{0}
\renewcommand{\theequation}{S\arabic{equation}}
\setcounter{equation}{0}

\section{Supplementary Information}
\subsection{Experimental setup}

We generate pairs of spatially entangled photons via type-I spontaneous parametric down conversion (SPDC) in an $8 mm$ long BBO crystal. The crystal is pumped with a classical $405 nm$ continues wave laser with a power of roughly $100mW$ and a waist of $w\approx600\mu m$. After a $150 mm$ lens, the photons pass through a binary amplitude mask, where two rows of circular apertures with a radius of $100\mu m$ and a horizontal spacing of $300\mu m$ carve five spatial modes per photon. A vertical distance of $1.5mm$ separates the two rows. Two sets of 4f lens systems image the spatial modes onto the first plane of a 10-plane light converter. The light converter consists of an SLM (Hamamtsu X13138-02), a dielectric mirror, and a prism. The photons bounce between the top half of the SLM and the mirror five times before being reflected back by the prism, bouncing five additional times on the bottom half. The distance between the SLM and the mirror is $43.5 mm$, and between the SLM and the prism is $69 mm$. Each phase mask on the SLM has a size of 140-by-360 pixels. Two motorized $100\mu m$ multimode fibers coupled to single-photon detectors positioned at the ‘11th’ plane ($87 mm$ after the 10th plane) are used to measure the correlations between the two photons at the output of the light converter, with a coincidence window of $400 ps$. A polarizer and $20nm$ spectral filters are used to reduce the effect of background and unmodulated light. Further details on the alignment and characterization of the experimental setup, which are beyond the focus of this work, can be found in refs. \cite{lib2024resource,lib2024high,lib2024building}.

\subsection{Quantum state tomography}

For $d=5$, we consider mutually unbiased bases (MUBs) denoted by $\left|m_{\alpha, i}\right\rangle$, where $\alpha$ ranges from $1$ to $d+1$. Specifically, $\alpha=1$ corresponds to the computational basis, while $\alpha=2$ to $d+1$ represent the other MUBs. Within each MUB, the index $i$ ranges from 1 to $d$. We project the quantum state $\varrho_{A B}$ onto $\left|m_{\alpha, i}\right\rangle_A \otimes\left|m_{\beta, j}\right\rangle_B^*$. The probability of detecting the coincident photons in the specific basis $\left|m_{\alpha, i}\right\rangle_A$ and $\left|m_{\beta, j}\right\rangle_B^*$ is given by
\begin{equation}
p_{\alpha,\beta,i,j}^{(M)}=\left\langle m_{\alpha,i}\right|_A\left\langle m_{\beta,j}\right|_B^* \varrho_{AB} \left|m_{\alpha,i}\right\rangle_A\left|m_{\beta,j}\right\rangle_B^*.
\end{equation}
They follow the normalization condition $\sum_{i,j}p_{\alpha,\beta,i,j}^{(M)}=1$ for any given $\alpha$ and $\beta$. After collecting photon counts from MUB measurements, we can calculate the probability distribution $p_{\alpha,\beta,i,j}$ which provide overcomplete tomography data. We follow the reconstruction strategy detailed in~\cite{GiovanniniCharacterizationPRL2013}, adapted to our scenario. The reconstruction proceeds as follows. First, we construct a guessed density matrix $\varrho_g$ using
\begin{equation}
\varrho_g=\frac{G^{\dagger} G}{\operatorname{Tr}\left(G^{\dagger} G\right)},
\end{equation}
where $G$ is defined as a linear combination of the identity matrix and the Gell-Mann matrices $g_k$ for dimension $D=d^2$,
\begin{equation}
G=c_0 \mathbb{1} +\sum_{k=1}^{D^2-1} c_k g_k.
\end{equation}
This construction ensures that all eigenvalues of $\varrho_g$ are nonnegative~\cite{AgnewTomographyPRA2011}. Next, we perform a numerical optimization to minimize the Chi-square quantity~\cite{OpatrnyLeastPRA1997,BanaszekMaximumPRA1999}
\begin{equation}
\chi^2=\sum_{\alpha,\beta,i,j} \frac{\left(p_{\alpha,\beta,i,j}^{(M)}-p_{\alpha,\beta,i,j}^{(P)}\right)^2}{p_{\alpha,\beta,i,j}^{(P)}}.
\end{equation}
where $p_i^{(M)}$ are the probabilities computed from the experiment data, and $p_{\alpha,\beta,i,j}^{(P)}$ are the predicted probabilities calculated from the guessed density matrix \begin{equation}
p_{\alpha,\beta,i,j}^{(P)}=\left\langle m_{\alpha,i}\right|_A\left\langle m_{\beta,j}\right|_B^* \varrho_{g} \left|m_{\alpha,i}\right\rangle_A\left|m_{\beta,j}\right\rangle_B^*.
\end{equation}

\begin{figure}[ht!]
\centering
\includegraphics[width=0.7\columnwidth]{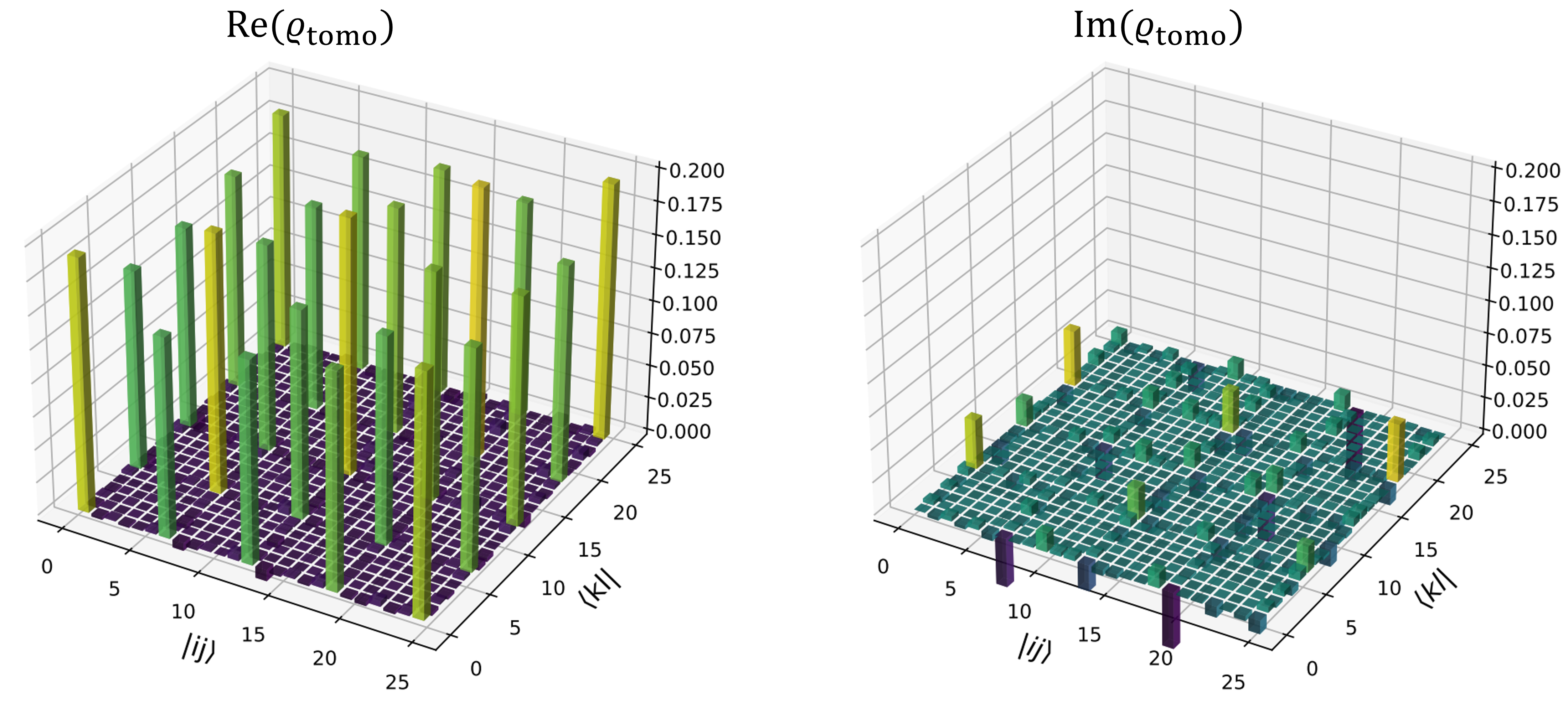}
\caption{The tomography result of $\varrho_{\text{tomo}}$. The axis ticks represent the order of the state basis $\ket{ij}$ and $\bra{kl}$. The left panel shows the real part, while the right panel shows the imaginary part.}
\label{figS:rhoTomoWeighted}
\end{figure}

In our experimental setup, measurements in the computational basis produce much higher photon counts than those in the other MUBs, making them more reliable. To account for the varying photon counts during numerical optimization, we incorporate them by assigning extra weights to each term in the chi-squared quantity. This weighting can be effectively interpreted as additional measurements. For instance, if the photon counts for measurements in the computational basis are ten times those of the other MUBs, we treat a single measurement in the computational basis as ten separate measurements. This approach equalizes the total photon counts for each measurement setting $\alpha,\beta$, transforming high photon counts into more repeated measurements and low photon counts into fewer repetitions. Therefore, we perform another numerical optimization to minimize the following weighted quantity,
\begin{equation}
\sum_{\alpha, \beta, i,j} \frac{\left(p_{\alpha, \beta, i,j}^{(M)}-p_{\alpha, \beta, i,j}^{(P)}\right)^2}{p_{\alpha, \beta, i,j}^{(P)}} N_{\alpha, \beta}^{(M)}.
\end{equation}
where $N_{\alpha, \beta}^{(M)}=\sum_{i,j}C_{\alpha, \beta, i,j}^{(M)}$ is the total photon count for the measurement setting $(\alpha, \beta)$ across all $i,j$. By weighting the terms to reflect the reliability of measurements in the computational basis, we observe that the tomography fits the experimental data more accurately. See \cref{figS:rhoTomoWeighted} for $\varrho_{\text{tomo}}$.

\subsection{Schmidt number certification method}

One important aspect of this method relies on the fact that for specific choices of observables $M_a,M_b$, this randomized correlator $x_i$ can be used to estimate the $su(d)$ Bloch sphere integrals defined by 
\begin{equation}
\mathcal{S}_{\varrho}^{(m)}:=N \int d \hat{\boldsymbol{\alpha}} \int d \hat{\boldsymbol{\beta}}\left[\av{g_{\hat{\boldsymbol{\alpha}}} \otimes g_{\hat{\boldsymbol{\beta}}}}_{\varrho}\right]^m ,
\end{equation}
where $g_{\hat{\boldsymbol{\alpha}}}:=\hat{\boldsymbol{\alpha}} \cdot \boldsymbol{g}$ and $g_{\hat{\boldsymbol{\beta}}}:=\hat{\boldsymbol{\beta}} \cdot \boldsymbol{g}$ are $su(d)$ operators determined by two real unit vectors $\boldsymbol{\alpha},\boldsymbol{\beta}$ in the $(d^2-1)$-dimensional Bloch sphere~\cite{BoundImaiPRL2021}. 
Those are in turn related to randomized correlators of the form
\begin{equation}
\mathfrak{R}_{\varrho}^{(t)}=\int {\mathrm d} U_a {\mathrm d} U_b \av{\left(U_a \otimes U_b\right) M \otimes M\left(U_a \otimes U_b\right)^{\dagger}}^{t} ,
\end{equation}
where $M$ is a given observable and
$U_a$ and $U_b$ are random unitaries sampled from the Haar measure ${\mathrm d} U$. 

The estimation of the second- and fourth-order integrals $\left(\mathcal{S}_{\varrho}^{(2)}, \mathcal{S}_{\varrho}^{(4)}\right)$ proceeds as  follows~\cite{liu2023characterizing}. First, we estimate the mean value as
\begin{equation}
\begin{aligned}
\bar{\mathcal{S}}_{\varrho}^{(2)}&=(d+1)^2 \bar{\mathfrak{R}}_{\varrho}^{(2)}= (d+1)^2 \frac{1}{N_{\mathrm{tot}}} \sum_i x_i^2\left(U_a \otimes U_b\right)  , \\
\bar{\mathcal{S}}_{\varrho}^{(4)}&=\frac{(d+1)^2\left(d^2+1\right)^2}{9(d-1)^2} \bar{\mathfrak{R}}_{\varrho}^{(4)} = \frac{(d+1)^2\left(d^2+1\right)^2}{9(d-1)^2} \frac{1}{N_{\mathrm{tot}}} \sum_i x_i^4\left(U_a \otimes U_b\right) ,
\end{aligned}
\end{equation}
where in the last equality we expressed the fact that the mean values of the randomized correlators are given by
\begin{equation}
\bar{\mathfrak{R}}_{\varrho}^{(t)}= \frac{1}{N_{\mathrm{tot}}} \sum_{i=1}^{N_{\mathrm{tot}}}\av{\left(U_a \otimes U_b\right)_i M \otimes M\left(U_a \otimes U_b\right)_i^{\dagger}}^{t} := \frac{1}{N_{\mathrm{tot}}} \sum_i x_i^t\left(U_a \otimes U_b\right) ,
\end{equation}
where $i$ labels the sample points.

\subsection{The effect of finite statistics}\label{sec:FiniteStatistics}

\begin{figure}[ht!]
\centering
\includegraphics[width=0.7\columnwidth]{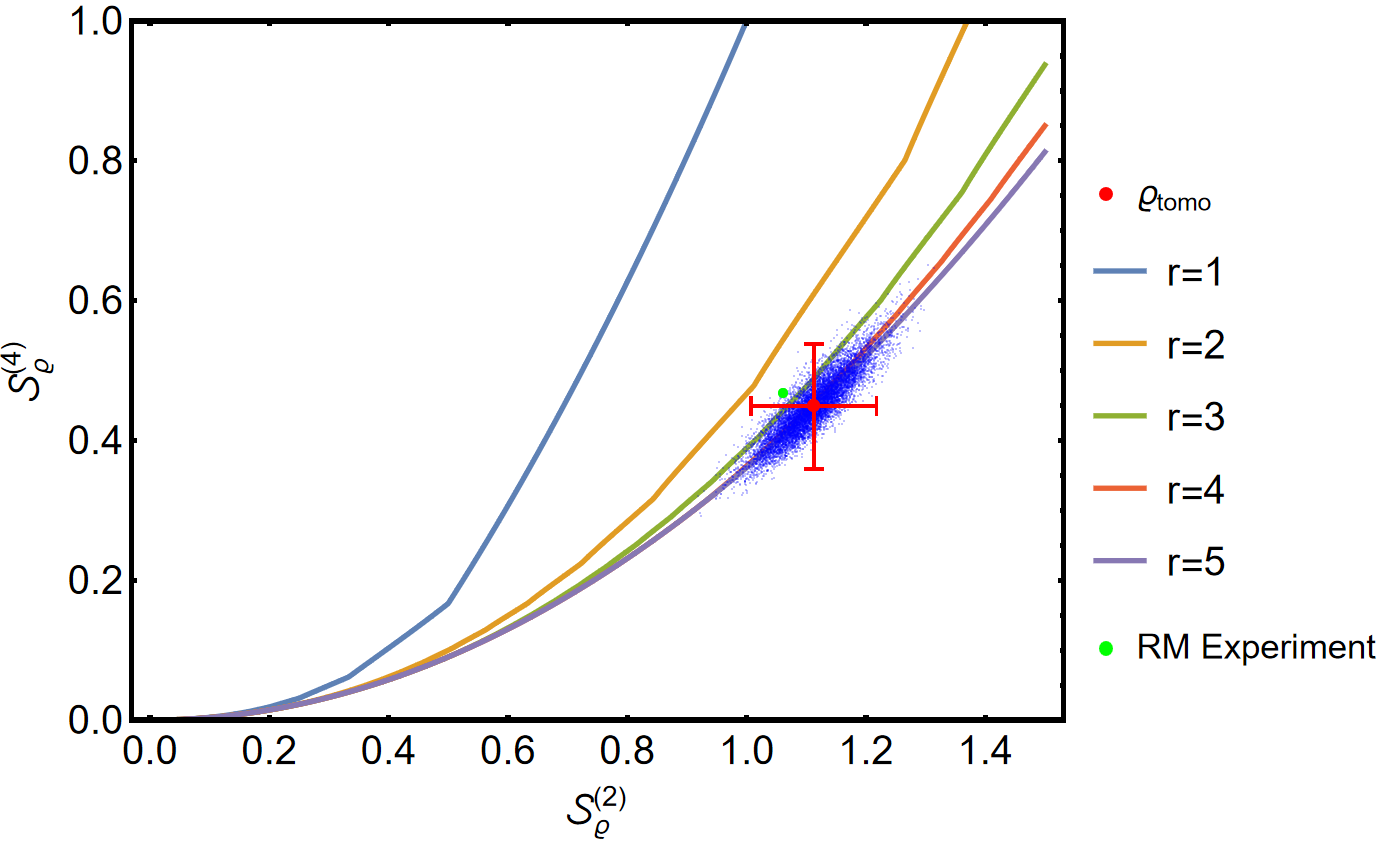}
\caption{The solid piecewise lines are the boundaries for all Schmidt numbers. The red point represents the reconstructed state with simulated measurements. The error bars are the $2\sigma$ uncertainty region. The green point corresponds to the result of the randomized measurement.}
\label{figS:FiniteStatisticsUncertainty}
\end{figure}

To examine the effect of finite statistics, we fix the total number of unitary transformations to $N_{\mathrm{tot}}=800$. Using the quantum state reconstructed from the tomography process, we compute the mean values 
\begin{equation}
    \begin{aligned}
        \bar{\mathcal{S}}_{\varrho}^{(2)} &= 1.11 , \\
        \bar{\mathcal{S}}_{\varrho}^{(4)}  &= 0.45 , 
    \end{aligned}
\end{equation}
as shown in the plot below. Next, we calculate the standard deviations resulting from the finite number of unitaries. Employing the analytical relations from~\cite{liu2023characterizing}, the variances are given by
\begin{equation}
\begin{aligned}
\sigma^2\left[\mathcal{S}_{\varrho}^{(2)}\right] & =\frac{(d+1)^4}{N_{\mathrm{tot}}}\left(\bar{\mathfrak{R}}_{\varrho}^{(4)}-\left(\bar{\mathfrak{R}}_{\varrho}^{(2)}\right)^2\right) \\
\sigma^2\left[\mathcal{S}_{\varrho}^{(4)}\right] & =\frac{(d+1)^4\left(d^2+1\right)^4}{81(d-1)^4 N_{\mathrm{tot}}}\left(\bar{\mathfrak{R}}_{\varrho}^{(8)}-\left(\bar{\mathfrak{R}}_{\varrho}^{(4)}\right)^2\right) .
\end{aligned}
\end{equation}
We generate $8\times 10^{6}$ unitaries to estimate $\bar{\mathfrak{R}}_{\varrho}^{(t)}$ for $t=2,4,8$, obtaining
\begin{equation}
\begin{aligned}
\bar{\mathfrak{R}}_{\varrho}^{(2)} &\approx 0.0309\\
\bar{\mathfrak{R}}_{\varrho}^{(4)} &\approx 0.0027\\
\bar{\mathfrak{R}}_{\varrho}^{(8)} &\approx 0.000062
\end{aligned}
\end{equation}
Consequently, we have
\begin{equation}
\begin{aligned}
\sigma^2\left[\mathcal{S}_{\varrho}^{(2)}\right] &\approx 0.0028\\
\sigma^2\left[\mathcal{S}_{\varrho}^{(4)}\right] &\approx 0.0020\\
\end{aligned}
\end{equation}
In \cref{figS:FiniteStatisticsUncertainty} we plot a $2\sigma$ uncertainty region around the point representing the ideal result of the reconstructed state.

To simulate multiple finite-statistics outcomes, we partition the unitaries into 10000 sets, each containing 800 unitaries. The resulting $\left(\mathcal{S}_{\varrho}^{(2)}, \mathcal{S}_{\varrho}^{(4)}\right)$ from these sets are plotted as blue dots in \cref{figS:FiniteStatisticsUncertainty}.

\subsection{Effect of random phase noise: entanglement detection using two MUBs vs randomized measurements}

The mode-dependent phase noise can be represented by the operator
$D=\sum_{i=1}^d e^{i \phi_i(t)}|i\rangle\langle i|$. Before applying the local random unitaries, the state experiences this random phase noise. Therefore, the randomized correlator that we measure is affected by the random phase noise as follows
\begin{equation}
x_i^{\prime}\left(U_a \otimes U_b\right):=\left\langle D_a U_a M_a U_a^{\dagger} D_a^{\dagger} \otimes D_b U_b M_b U_b^{\dagger} D_b^{\dagger}\right\rangle.
\end{equation}
To investigate the effect of the random phases, we add different random phases to each of the 800 unitaries implemented in the experiment and calculate a corresponding point in the $\left(\mathcal{S}_{\varrho}^{(2)}, \mathcal{S}_{\varrho}^{(4)}\right)$ plane. Repeating this process $10^6$ times yields a distribution in the plane. This distribution is plotted in \cref{fig:3} in the main text, where regions of high probability are shown in yellow, while regions of low probability are shown in purple. As can be seen from the plot, the most probable region lies almost entirely within the area corresponding to Schmidt number $r\geq 4$, indicating that our method is quite robust against mode-dependent phase noise.

As a comparison, we can consider an alternative simple and powerful method that can be used for detecting the entanglement dimensionality in experiments like ours, that is to consider measurements in two MUBs, of which canonical examples are two bases connected via Discrete Fourier Transform (DFT)~\cite{namiki2012discrete,ErkerQuantum2017quantifyinghigh,BavarescoMeasurementsNP2018,HerreraHighQuantum2020,valencia2020unscrambling,ResourceMorelliPRL2023}.  
Thus, we consider the Fourier transform of the computational basis, denoted by $\ket{\bar{j}}$ which satisfy the periodic condition $\ket{\overline{j+d}}=\ket{\bar{j}}$:
\begin{equation}
\ket{\bar{j}}=F \ket{j} = \frac{1}{\sqrt{d}} \sum_{k=0}^{d-1} \omega^{jk} \ket{k} ,  
\end{equation}
where 
\begin{equation}
    F= \frac{1}{\sqrt{d}} \sum_{jk} \omega^{jk} \ketbra{k}{j} 
\end{equation}
is the Discrete Fourier Transform and $\omega=e^{i(2 \pi / d)}$. 
Using both the computational basis and the DFT basis, a canonical Schmidt number witness can be constructed from the correlator~\cite{namiki2012discrete} 
\begin{equation}
C_d:=\sum_{j=0}^{d-1}(|j\rangle\langle j| \otimes|j\rangle\langle j|+|\bar{j}\rangle\langle\bar{j}| \otimes|\overline{-j}\rangle\langle\overline{-j}|)
\end{equation}
Concretely, whenever a state $\varrho$ satisfies
\begin{equation}
\begin{aligned}\label{eq;DFTwitness}
\av{C_d}_\varrho & > 1+\frac{r-1}{d} , 
\end{aligned}
\end{equation}
this implies that its Schmidt number is at least $r$.
Similar methods to detect the Schmidt number from two or more mutually unbiased bases exist~\cite{namiki2012discrete,ResourceMorelliPRL2023}, but they lead to similar conclusions in our case so we do not present their details here for simplicity.

\begin{figure}[ht!]
\centering
\includegraphics[width=0.6\columnwidth]{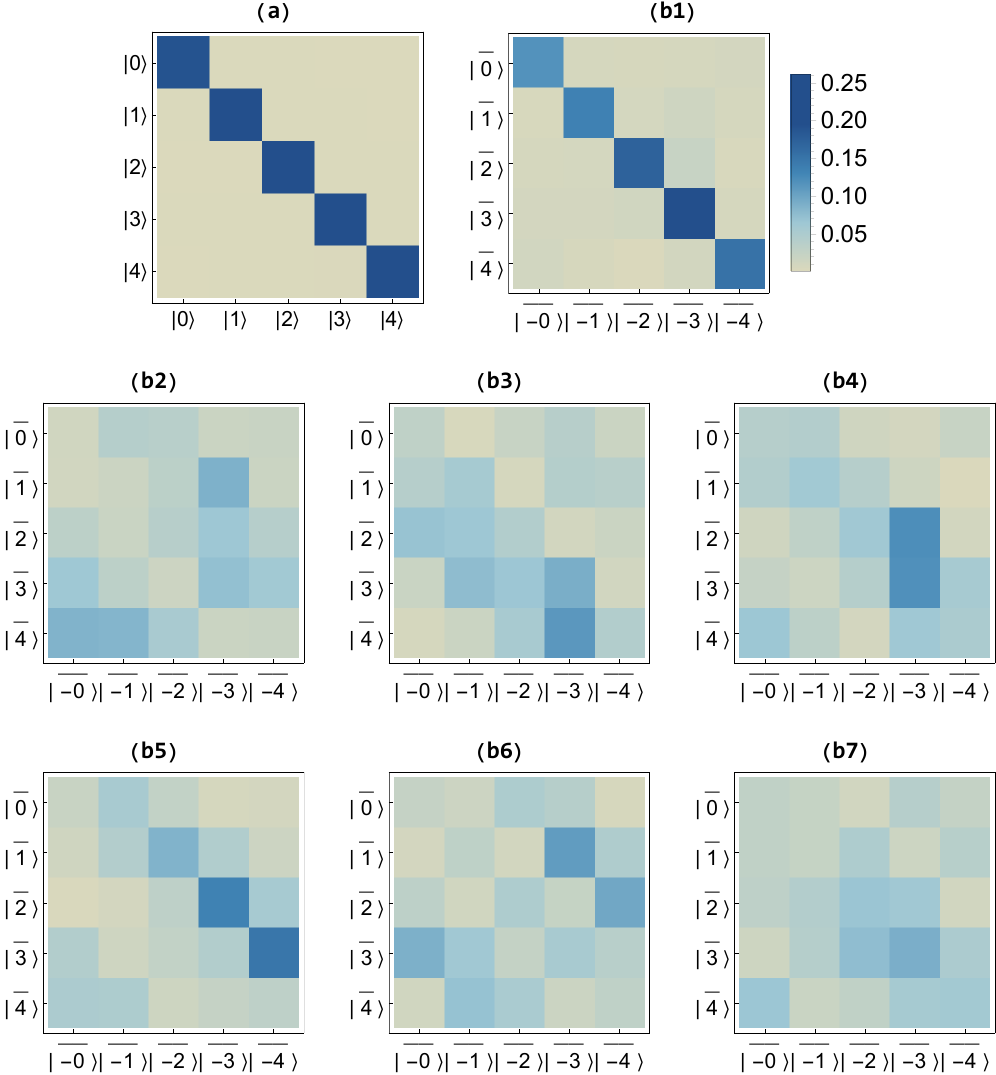}
\caption{(a) The measured probabilities in the computational basis $\ket{j,k}$. (b1)-(b7) The measured probabilities in the DFT basis $\ket{\overline{j},\overline{k}}$. In (a) and (b1), the state is measured without adding extra mode-dependent random phase noise, so the plots exhibit a diagonal feature. In contrast, in (b2)-(b7) we added such random noise.}
\label{figS:DFTData}
\end{figure}

Our measured probabilities in these two bases are plotted in \cref{figS:DFTData}. 
By combining (a) with each of (b1)–(b7), we can evaluate the criterion in \cref{eq;DFTwitness}. The results are summarized in \cref{table:randomphases} and show that without adding mode-dependent random phase noise (b1) the detected Schmidt number is $r=5$. However, in the other six situations where random phases are added, the detected Schmidt numbers are reduced to $r=1, 2, 2, 1, 1, 2$ respectively. This demonstrates that the method is highly sensitive to such phase noise.

\begin{table}[H]
\centering
\begin{tabular}{|c|c|c|}
  \hline & \av{C_d} & \text{Schmidt number} \\
  \hline
  \text{No random phase} & 1.83184 & 5 \\
  \hline
  \text{random phase 1} & 1.15041 & 1 \\
  \hline
  \text{random phase 2} & 1.25493 & 2 \\
  \hline
  \text{random phase 3} & 1.32341 & 2 \\
  \hline
  \text{random phase 4} & 1.15997 & 1 \\
  \hline
  \text{random phase 5} & 1.17687 & 1 \\
  \hline
  \text{random phase 6} & 1.25993 & 2 \\
  \hline
\end{tabular}
\caption{Schmidt number witness based on computational and DFT bases with random phase noise.}\label{table:randomphases}
\end{table}

\ignore{
\begin{equation}
\begin{array}{|c|c|c|c|c|}
  \hline
  & \av{C_d} & r \text{ given by } \langle \hat{C}_d
  \rangle & \langle \hat{R}_d \rangle & r \text{ given by } \langle
  \hat{R}_d \rangle\\
  \hline
  \text{No random phase} & 1.83184 & 5 & 1.80744 & 4\\
  \hline
  \text{random phase 1} & 1.15041 & 1 & 0.88386 & 1\\
  \hline
  \text{random phase 2} & 1.25493 & 2 & 1.03860 & 1\\
  \hline
  \text{random phase 3} & 1.32341 & 2 & 1.37025 & 1\\
  \hline
  \text{random phase 4} & 1.15997 & 1 & 1.10667 & 1\\
  \hline
  \text{random phase 5} & 1.17687 & 1 & 0.70145 & 1\\
  \hline
  \text{random phase 6} & 1.25993 & 2 & 1.20177 & 1\\
  \hline
\end{array}
\end{equation}
}

\subsection{The optimal dephasing model approximation}
To explain both the moments of the measured random correlations (the 800 chosen unitaries) and the reconstructed density matrix, we use a dephasing model based on fluctuating phases between different spatial modes. For this, we calculate the following density matrix,
\begin{equation}
\varrho_{\mathrm{deph}}=\sum_{i=1}^{n} \frac{1}{n} \left|\psi_i\right\rangle\left\langle\psi_i\right| \text { with }\left|\psi_i\right\rangle=\sum_{k=1}^{5} e^{i \phi_{i, k}^{(a)}}|k\rangle \otimes e^{i \phi_{i, k}^{(b)}}|k^{\prime}\rangle,
\end{equation}
where $n$ is a sufficiently large number, such as $10^6$. Here, each $\phi_{i, k}^{(a / b)}$ is a random phase uniformly distributed within $-\phi_{\max} \leq \phi_{i, k}^{(a / b)} \leq \phi_{\max}$. 
We find that when $\phi_{\max}=0.53$, the point on the $(\bar{\mathcal{S}}_{\varrho}^{(2)} , \bar{\mathcal{S}}_{\varrho}^{(4)})$ plane corresponding to a simulation using $\varrho_{\mathrm{deph}}$ and the same 800 chosen unitaries as in the main text is closest to the experimental data from randomized measurements, with a distance of $\sqrt{\Delta^2 \bar{\mathcal{S}}_{\varrho}^{(2)}+\Delta^2 \bar{\mathcal{S}}_{\varrho}^{(4)}}\approx 0.02$. This distance is of the same magnitude as two standard deviations in \cref{fig:2} of the main text. The result indicates that the best approximation is achieved when $\phi_{\max}=0.53$.

%% file: main.bbl
\begin{thebibliography}{61}%
\makeatletter
\providecommand \@ifxundefined [1]{%
 \@ifx{#1\undefined}
}%
\providecommand \@ifnum [1]{%
 \ifnum #1\expandafter \@firstoftwo
 \else \expandafter \@secondoftwo
 \fi
}%
\providecommand \@ifx [1]{%
 \ifx #1\expandafter \@firstoftwo
 \else \expandafter \@secondoftwo
 \fi
}%
\providecommand \natexlab [1]{#1}%
\providecommand \enquote  [1]{``#1''}%
\providecommand \bibnamefont  [1]{#1}%
\providecommand \bibfnamefont [1]{#1}%
\providecommand \citenamefont [1]{#1}%
\providecommand \href@noop [0]{\@secondoftwo}%
\providecommand \href [0]{\begingroup \@sanitize@url \@href}%
\providecommand \@href[1]{\@@startlink{#1}\@@href}%
\providecommand \@@href[1]{\endgroup#1\@@endlink}%
\providecommand \@sanitize@url [0]{\catcode `\\12\catcode `\$12\catcode `\&12\catcode `\#12\catcode `\^12\catcode `\_12\catcode `\%12\relax}%
\providecommand \@@startlink[1]{}%
\providecommand \@@endlink[0]{}%
\providecommand \url  [0]{\begingroup\@sanitize@url \@url }%
\providecommand \@url [1]{\endgroup\@href {#1}{\urlprefix }}%
\providecommand \urlprefix  [0]{URL }%
\providecommand \Eprint [0]{\href }%
\providecommand \doibase [0]{https://doi.org/}%
\providecommand \selectlanguage [0]{\@gobble}%
\providecommand \bibinfo  [0]{\@secondoftwo}%
\providecommand \bibfield  [0]{\@secondoftwo}%
\providecommand \translation [1]{[#1]}%
\providecommand \BibitemOpen [0]{}%
\providecommand \bibitemStop [0]{}%
\providecommand \bibitemNoStop [0]{.\EOS\space}%
\providecommand \EOS [0]{\spacefactor3000\relax}%
\providecommand \BibitemShut  [1]{\csname bibitem#1\endcsname}%
\let\auto@bib@innerbib\@empty
\bibitem [{\citenamefont {Freedman}\ and\ \citenamefont {Clauser}(1972)}]{freedman1972experimental}%
  \BibitemOpen
  \bibfield  {author} {\bibinfo {author} {\bibfnamefont {S.~J.}\ \bibnamefont {Freedman}}\ and\ \bibinfo {author} {\bibfnamefont {J.~F.}\ \bibnamefont {Clauser}},\ }\href {https://doi.org/10.1103/PhysRevLett.28.938} {\bibfield  {journal} {\bibinfo  {journal} {Phys. Rev. Lett.}\ }\textbf {\bibinfo {volume} {28}},\ \bibinfo {pages} {938} (\bibinfo {year} {1972})}\BibitemShut {NoStop}%
\bibitem [{\citenamefont {Aspect}\ \emph {et~al.}(1982)\citenamefont {Aspect}, \citenamefont {Dalibard},\ and\ \citenamefont {Roger}}]{aspect1982experimental}%
  \BibitemOpen
  \bibfield  {author} {\bibinfo {author} {\bibfnamefont {A.}~\bibnamefont {Aspect}}, \bibinfo {author} {\bibfnamefont {J.}~\bibnamefont {Dalibard}},\ and\ \bibinfo {author} {\bibfnamefont {G.}~\bibnamefont {Roger}},\ }\href {https://doi.org/10.1103/PhysRevLett.49.1804} {\bibfield  {journal} {\bibinfo  {journal} {Phys. Rev. Lett.}\ }\textbf {\bibinfo {volume} {49}},\ \bibinfo {pages} {1804} (\bibinfo {year} {1982})}\BibitemShut {NoStop}%
\bibitem [{\citenamefont {Jozsa}\ and\ \citenamefont {Linden}(2003)}]{jozsa2003role}%
  \BibitemOpen
  \bibfield  {author} {\bibinfo {author} {\bibfnamefont {R.}~\bibnamefont {Jozsa}}\ and\ \bibinfo {author} {\bibfnamefont {N.}~\bibnamefont {Linden}},\ }\href {https://doi.org/https://doi.org/10.1098/rspa.2002.1097} {\bibfield  {journal} {\bibinfo  {journal} {Proc. R. Soc. Lond. A.}\ }\textbf {\bibinfo {volume} {459}},\ \bibinfo {pages} {2011} (\bibinfo {year} {2003})}\BibitemShut {NoStop}%
\bibitem [{\citenamefont {Ekert}(1991)}]{ekert1991quantum}%
  \BibitemOpen
  \bibfield  {author} {\bibinfo {author} {\bibfnamefont {A.~K.}\ \bibnamefont {Ekert}},\ }\href {https://doi.org/10.1103/PhysRevLett.67.661} {\bibfield  {journal} {\bibinfo  {journal} {Phys. Rev. Lett.}\ }\textbf {\bibinfo {volume} {67}},\ \bibinfo {pages} {661} (\bibinfo {year} {1991})}\BibitemShut {NoStop}%
\bibitem [{\citenamefont {G{\"u}hne}\ and\ \citenamefont {T{\'o}th}(2009)}]{guhne2009entanglement}%
  \BibitemOpen
  \bibfield  {author} {\bibinfo {author} {\bibfnamefont {O.}~\bibnamefont {G{\"u}hne}}\ and\ \bibinfo {author} {\bibfnamefont {G.}~\bibnamefont {T{\'o}th}},\ }\href {https://doi.org/https://doi.org/10.1016/j.physrep.2009.02.004} {\bibfield  {journal} {\bibinfo  {journal} {Phys. Rep.}\ }\textbf {\bibinfo {volume} {474}},\ \bibinfo {pages} {1} (\bibinfo {year} {2009})}\BibitemShut {NoStop}%
\bibitem [{\citenamefont {Friis}\ \emph {et~al.}(2019)\citenamefont {Friis}, \citenamefont {Vitagliano}, \citenamefont {Malik},\ and\ \citenamefont {Huber}}]{friis2019entanglement}%
  \BibitemOpen
  \bibfield  {author} {\bibinfo {author} {\bibfnamefont {N.}~\bibnamefont {Friis}}, \bibinfo {author} {\bibfnamefont {G.}~\bibnamefont {Vitagliano}}, \bibinfo {author} {\bibfnamefont {M.}~\bibnamefont {Malik}},\ and\ \bibinfo {author} {\bibfnamefont {M.}~\bibnamefont {Huber}},\ }\href {https://doi.org/https://doi.org/10.1038/s42254-018-0003-5} {\bibfield  {journal} {\bibinfo  {journal} {Nat. Rev. Phys.}\ }\textbf {\bibinfo {volume} {1}},\ \bibinfo {pages} {72} (\bibinfo {year} {2019})}\BibitemShut {NoStop}%
\bibitem [{\citenamefont {Erhard}\ \emph {et~al.}(2020)\citenamefont {Erhard}, \citenamefont {Krenn},\ and\ \citenamefont {Zeilinger}}]{erhard2020advances}%
  \BibitemOpen
  \bibfield  {author} {\bibinfo {author} {\bibfnamefont {M.}~\bibnamefont {Erhard}}, \bibinfo {author} {\bibfnamefont {M.}~\bibnamefont {Krenn}},\ and\ \bibinfo {author} {\bibfnamefont {A.}~\bibnamefont {Zeilinger}},\ }\href {https://doi.org/https://doi.org/10.1038/s42254-020-0193-5} {\bibfield  {journal} {\bibinfo  {journal} {Nat. Rev. Phys.}\ }\textbf {\bibinfo {volume} {2}},\ \bibinfo {pages} {365} (\bibinfo {year} {2020})}\BibitemShut {NoStop}%
\bibitem [{\citenamefont {Terhal}\ and\ \citenamefont {Horodecki}(2000)}]{SchmidtBarbaraPRA2000}%
  \BibitemOpen
  \bibfield  {author} {\bibinfo {author} {\bibfnamefont {B.~M.}\ \bibnamefont {Terhal}}\ and\ \bibinfo {author} {\bibfnamefont {P.}~\bibnamefont {Horodecki}},\ }\href {https://doi.org/10.1103/PhysRevA.61.040301} {\bibfield  {journal} {\bibinfo  {journal} {Phys. Rev. A}\ }\textbf {\bibinfo {volume} {61}},\ \bibinfo {pages} {040301} (\bibinfo {year} {2000})}\BibitemShut {NoStop}%
\bibitem [{\citenamefont {Erker}\ \emph {et~al.}(2017)\citenamefont {Erker}, \citenamefont {Krenn},\ and\ \citenamefont {Huber}}]{ErkerQuantum2017quantifyinghigh}%
  \BibitemOpen
  \bibfield  {author} {\bibinfo {author} {\bibfnamefont {P.}~\bibnamefont {Erker}}, \bibinfo {author} {\bibfnamefont {M.}~\bibnamefont {Krenn}},\ and\ \bibinfo {author} {\bibfnamefont {M.}~\bibnamefont {Huber}},\ }\href {https://doi.org/10.22331/q-2017-07-28-22} {\bibfield  {journal} {\bibinfo  {journal} {{Quantum}}\ }\textbf {\bibinfo {volume} {1}},\ \bibinfo {pages} {22} (\bibinfo {year} {2017})}\BibitemShut {NoStop}%
\bibitem [{\citenamefont {Huber}\ \emph {et~al.}(2013)\citenamefont {Huber}, \citenamefont {Perarnau-Llobet},\ and\ \citenamefont {de~Vicente}}]{HuberEntropyPRA2013}%
  \BibitemOpen
  \bibfield  {author} {\bibinfo {author} {\bibfnamefont {M.}~\bibnamefont {Huber}}, \bibinfo {author} {\bibfnamefont {M.}~\bibnamefont {Perarnau-Llobet}},\ and\ \bibinfo {author} {\bibfnamefont {J.~I.}\ \bibnamefont {de~Vicente}},\ }\href {https://doi.org/10.1103/PhysRevA.88.042328} {\bibfield  {journal} {\bibinfo  {journal} {Phys. Rev. A}\ }\textbf {\bibinfo {volume} {88}},\ \bibinfo {pages} {042328} (\bibinfo {year} {2013})}\BibitemShut {NoStop}%
\bibitem [{\citenamefont {Huber}\ and\ \citenamefont {de~Vicente}(2013)}]{HuberStructurePRL2013}%
  \BibitemOpen
  \bibfield  {author} {\bibinfo {author} {\bibfnamefont {M.}~\bibnamefont {Huber}}\ and\ \bibinfo {author} {\bibfnamefont {J.~I.}\ \bibnamefont {de~Vicente}},\ }\href {https://doi.org/10.1103/PhysRevLett.110.030501} {\bibfield  {journal} {\bibinfo  {journal} {Phys. Rev. Lett.}\ }\textbf {\bibinfo {volume} {110}},\ \bibinfo {pages} {030501} (\bibinfo {year} {2013})}\BibitemShut {NoStop}%
\bibitem [{\citenamefont {Liu}\ \emph {et~al.}(2024)\citenamefont {Liu}, \citenamefont {Fadel}, \citenamefont {He}, \citenamefont {Huber},\ and\ \citenamefont {Vitagliano}}]{Liu2024bounding}%
  \BibitemOpen
  \bibfield  {author} {\bibinfo {author} {\bibfnamefont {S.}~\bibnamefont {Liu}}, \bibinfo {author} {\bibfnamefont {M.}~\bibnamefont {Fadel}}, \bibinfo {author} {\bibfnamefont {Q.}~\bibnamefont {He}}, \bibinfo {author} {\bibfnamefont {M.}~\bibnamefont {Huber}},\ and\ \bibinfo {author} {\bibfnamefont {G.}~\bibnamefont {Vitagliano}},\ }\href {https://doi.org/10.22331/q-2024-01-30-1236} {\bibfield  {journal} {\bibinfo  {journal} {{Quantum}}\ }\textbf {\bibinfo {volume} {8}},\ \bibinfo {pages} {1236} (\bibinfo {year} {2024})}\BibitemShut {NoStop}%
\bibitem [{\citenamefont {Morelli}\ \emph {et~al.}(2023)\citenamefont {Morelli}, \citenamefont {Huber},\ and\ \citenamefont {Tavakoli}}]{ResourceMorelliPRL2023}%
  \BibitemOpen
  \bibfield  {author} {\bibinfo {author} {\bibfnamefont {S.}~\bibnamefont {Morelli}}, \bibinfo {author} {\bibfnamefont {M.}~\bibnamefont {Huber}},\ and\ \bibinfo {author} {\bibfnamefont {A.}~\bibnamefont {Tavakoli}},\ }\href {https://doi.org/10.1103/PhysRevLett.131.170201} {\bibfield  {journal} {\bibinfo  {journal} {Phys. Rev. Lett.}\ }\textbf {\bibinfo {volume} {131}},\ \bibinfo {pages} {170201} (\bibinfo {year} {2023})}\BibitemShut {NoStop}%
\bibitem [{\citenamefont {Ecker}\ \emph {et~al.}(2019)\citenamefont {Ecker}, \citenamefont {Bouchard}, \citenamefont {Bulla}, \citenamefont {Brandt}, \citenamefont {Kohout}, \citenamefont {Steinlechner}, \citenamefont {Fickler}, \citenamefont {Malik}, \citenamefont {Guryanova}, \citenamefont {Ursin},\ and\ \citenamefont {Huber}}]{EckerOvercomingPRX2019}%
  \BibitemOpen
  \bibfield  {author} {\bibinfo {author} {\bibfnamefont {S.}~\bibnamefont {Ecker}}, \bibinfo {author} {\bibfnamefont {F.}~\bibnamefont {Bouchard}}, \bibinfo {author} {\bibfnamefont {L.}~\bibnamefont {Bulla}}, \bibinfo {author} {\bibfnamefont {F.}~\bibnamefont {Brandt}}, \bibinfo {author} {\bibfnamefont {O.}~\bibnamefont {Kohout}}, \bibinfo {author} {\bibfnamefont {F.}~\bibnamefont {Steinlechner}}, \bibinfo {author} {\bibfnamefont {R.}~\bibnamefont {Fickler}}, \bibinfo {author} {\bibfnamefont {M.}~\bibnamefont {Malik}}, \bibinfo {author} {\bibfnamefont {Y.}~\bibnamefont {Guryanova}}, \bibinfo {author} {\bibfnamefont {R.}~\bibnamefont {Ursin}},\ and\ \bibinfo {author} {\bibfnamefont {M.}~\bibnamefont {Huber}},\ }\href {https://doi.org/10.1103/PhysRevX.9.041042} {\bibfield  {journal} {\bibinfo  {journal} {Phys. Rev. X}\ }\textbf {\bibinfo {volume} {9}},\ \bibinfo {pages} {041042} (\bibinfo {year} {2019})}\BibitemShut {NoStop}%
\bibitem [{\citenamefont {Hu}\ \emph {et~al.}(2021)\citenamefont {Hu}, \citenamefont {Zhang}, \citenamefont {Guo}, \citenamefont {Wang}, \citenamefont {Xing}, \citenamefont {Huang}, \citenamefont {Liu}, \citenamefont {Huang}, \citenamefont {Li}, \citenamefont {Guo}, \citenamefont {Gao}, \citenamefont {Pivoluska},\ and\ \citenamefont {Huber}}]{HuPathwaysPRL2021}%
  \BibitemOpen
  \bibfield  {author} {\bibinfo {author} {\bibfnamefont {X.-M.}\ \bibnamefont {Hu}}, \bibinfo {author} {\bibfnamefont {C.}~\bibnamefont {Zhang}}, \bibinfo {author} {\bibfnamefont {Y.}~\bibnamefont {Guo}}, \bibinfo {author} {\bibfnamefont {F.-X.}\ \bibnamefont {Wang}}, \bibinfo {author} {\bibfnamefont {W.-B.}\ \bibnamefont {Xing}}, \bibinfo {author} {\bibfnamefont {C.-X.}\ \bibnamefont {Huang}}, \bibinfo {author} {\bibfnamefont {B.-H.}\ \bibnamefont {Liu}}, \bibinfo {author} {\bibfnamefont {Y.-F.}\ \bibnamefont {Huang}}, \bibinfo {author} {\bibfnamefont {C.-F.}\ \bibnamefont {Li}}, \bibinfo {author} {\bibfnamefont {G.-C.}\ \bibnamefont {Guo}}, \bibinfo {author} {\bibfnamefont {X.}~\bibnamefont {Gao}}, \bibinfo {author} {\bibfnamefont {M.}~\bibnamefont {Pivoluska}},\ and\ \bibinfo {author} {\bibfnamefont {M.}~\bibnamefont {Huber}},\ }\href {https://doi.org/10.1103/PhysRevLett.127.110505} {\bibfield  {journal} {\bibinfo  {journal} {Phys. Rev. Lett.}\ }\textbf {\bibinfo {volume} {127}},\ \bibinfo {pages} {110505}
  (\bibinfo {year} {2021})}\BibitemShut {NoStop}%
\bibitem [{\citenamefont {Krenn}\ \emph {et~al.}(2014)\citenamefont {Krenn}, \citenamefont {Huber}, \citenamefont {Fickler}, \citenamefont {Lapkiewicz}, \citenamefont {Ramelow},\ and\ \citenamefont {Zeilinger}}]{KrennGenerationNAS2014}%
  \BibitemOpen
  \bibfield  {author} {\bibinfo {author} {\bibfnamefont {M.}~\bibnamefont {Krenn}}, \bibinfo {author} {\bibfnamefont {M.}~\bibnamefont {Huber}}, \bibinfo {author} {\bibfnamefont {R.}~\bibnamefont {Fickler}}, \bibinfo {author} {\bibfnamefont {R.}~\bibnamefont {Lapkiewicz}}, \bibinfo {author} {\bibfnamefont {S.}~\bibnamefont {Ramelow}},\ and\ \bibinfo {author} {\bibfnamefont {A.}~\bibnamefont {Zeilinger}},\ }\href {https://doi.org/10.1073/pnas.1402365111} {\bibfield  {journal} {\bibinfo  {journal} {Proc. Natl. Acad. Sci. U.S.A.}\ }\textbf {\bibinfo {volume} {111}},\ \bibinfo {pages} {6243} (\bibinfo {year} {2014})}\BibitemShut {NoStop}%
\bibitem [{\citenamefont {Bavaresco}\ \emph {et~al.}(2018)\citenamefont {Bavaresco}, \citenamefont {Herrera~Valencia}, \citenamefont {Kl{\"o}ckl}, \citenamefont {Pivoluska}, \citenamefont {Erker}, \citenamefont {Friis}, \citenamefont {Malik},\ and\ \citenamefont {Huber}}]{BavarescoMeasurementsNP2018}%
  \BibitemOpen
  \bibfield  {author} {\bibinfo {author} {\bibfnamefont {J.}~\bibnamefont {Bavaresco}}, \bibinfo {author} {\bibfnamefont {N.}~\bibnamefont {Herrera~Valencia}}, \bibinfo {author} {\bibfnamefont {C.}~\bibnamefont {Kl{\"o}ckl}}, \bibinfo {author} {\bibfnamefont {M.}~\bibnamefont {Pivoluska}}, \bibinfo {author} {\bibfnamefont {P.}~\bibnamefont {Erker}}, \bibinfo {author} {\bibfnamefont {N.}~\bibnamefont {Friis}}, \bibinfo {author} {\bibfnamefont {M.}~\bibnamefont {Malik}},\ and\ \bibinfo {author} {\bibfnamefont {M.}~\bibnamefont {Huber}},\ }\href {https://doi.org/https://doi.org/10.1038/s41567-018-0203-z} {\bibfield  {journal} {\bibinfo  {journal} {Nat. Phys.}\ }\textbf {\bibinfo {volume} {14}},\ \bibinfo {pages} {1032} (\bibinfo {year} {2018})}\BibitemShut {NoStop}%
\bibitem [{\citenamefont {Schneeloch}\ \emph {et~al.}(2019)\citenamefont {Schneeloch}, \citenamefont {Tison}, \citenamefont {Fanto}, \citenamefont {Alsing},\ and\ \citenamefont {Howland}}]{SchneelochQuantifyingNC2019}%
  \BibitemOpen
  \bibfield  {author} {\bibinfo {author} {\bibfnamefont {J.}~\bibnamefont {Schneeloch}}, \bibinfo {author} {\bibfnamefont {C.~C.}\ \bibnamefont {Tison}}, \bibinfo {author} {\bibfnamefont {M.~L.}\ \bibnamefont {Fanto}}, \bibinfo {author} {\bibfnamefont {P.~M.}\ \bibnamefont {Alsing}},\ and\ \bibinfo {author} {\bibfnamefont {G.~A.}\ \bibnamefont {Howland}},\ }\href {https://doi.org/10.1038/s41467-019-10810-z} {\bibfield  {journal} {\bibinfo  {journal} {Nat. Commun.}\ }\textbf {\bibinfo {volume} {10}},\ \bibinfo {pages} {2785} (\bibinfo {year} {2019})}\BibitemShut {NoStop}%
\bibitem [{\citenamefont {Herrera~Valencia}\ \emph {et~al.}(2020)\citenamefont {Herrera~Valencia}, \citenamefont {Srivastav}, \citenamefont {Pivoluska}, \citenamefont {Huber}, \citenamefont {Friis}, \citenamefont {McCutcheon},\ and\ \citenamefont {Malik}}]{HerreraHighQuantum2020}%
  \BibitemOpen
  \bibfield  {author} {\bibinfo {author} {\bibfnamefont {N.}~\bibnamefont {Herrera~Valencia}}, \bibinfo {author} {\bibfnamefont {V.}~\bibnamefont {Srivastav}}, \bibinfo {author} {\bibfnamefont {M.}~\bibnamefont {Pivoluska}}, \bibinfo {author} {\bibfnamefont {M.}~\bibnamefont {Huber}}, \bibinfo {author} {\bibfnamefont {N.}~\bibnamefont {Friis}}, \bibinfo {author} {\bibfnamefont {W.}~\bibnamefont {McCutcheon}},\ and\ \bibinfo {author} {\bibfnamefont {M.}~\bibnamefont {Malik}},\ }\href {https://doi.org/10.22331/q-2020-12-24-376} {\bibfield  {journal} {\bibinfo  {journal} {{Quantum}}\ }\textbf {\bibinfo {volume} {4}},\ \bibinfo {pages} {376} (\bibinfo {year} {2020})}\BibitemShut {NoStop}%
\bibitem [{\citenamefont {Huber}\ \emph {et~al.}(2018)\citenamefont {Huber}, \citenamefont {Lami}, \citenamefont {Lancien},\ and\ \citenamefont {M\"uller-Hermes}}]{HuberHighPRL2018}%
  \BibitemOpen
  \bibfield  {author} {\bibinfo {author} {\bibfnamefont {M.}~\bibnamefont {Huber}}, \bibinfo {author} {\bibfnamefont {L.}~\bibnamefont {Lami}}, \bibinfo {author} {\bibfnamefont {C.}~\bibnamefont {Lancien}},\ and\ \bibinfo {author} {\bibfnamefont {A.}~\bibnamefont {M\"uller-Hermes}},\ }\href {https://doi.org/10.1103/PhysRevLett.121.200503} {\bibfield  {journal} {\bibinfo  {journal} {Phys. Rev. Lett.}\ }\textbf {\bibinfo {volume} {121}},\ \bibinfo {pages} {200503} (\bibinfo {year} {2018})}\BibitemShut {NoStop}%
\bibitem [{\citenamefont {Kl\"ockl}\ and\ \citenamefont {Huber}(2015)}]{KlocklCharacterizingPRA2015}%
  \BibitemOpen
  \bibfield  {author} {\bibinfo {author} {\bibfnamefont {C.}~\bibnamefont {Kl\"ockl}}\ and\ \bibinfo {author} {\bibfnamefont {M.}~\bibnamefont {Huber}},\ }\href {https://doi.org/10.1103/PhysRevA.91.042339} {\bibfield  {journal} {\bibinfo  {journal} {Phys. Rev. A}\ }\textbf {\bibinfo {volume} {91}},\ \bibinfo {pages} {042339} (\bibinfo {year} {2015})}\BibitemShut {NoStop}%
\bibitem [{\citenamefont {Zhou}\ \emph {et~al.}(2020)\citenamefont {Zhou}, \citenamefont {Zeng},\ and\ \citenamefont {Liu}}]{Zhou2020}%
  \BibitemOpen
  \bibfield  {author} {\bibinfo {author} {\bibfnamefont {Y.}~\bibnamefont {Zhou}}, \bibinfo {author} {\bibfnamefont {P.}~\bibnamefont {Zeng}},\ and\ \bibinfo {author} {\bibfnamefont {Z.}~\bibnamefont {Liu}},\ }\href {https://link.aps.org/doi/10.1103/PhysRevLett.125.200502} {\bibfield  {journal} {\bibinfo  {journal} {Phys. Rev. Lett.}\ }\textbf {\bibinfo {volume} {125}},\ \bibinfo {pages} {200502} (\bibinfo {year} {2020})}\BibitemShut {NoStop}%
\bibitem [{\citenamefont {Elben}\ \emph {et~al.}(2020)\citenamefont {Elben}, \citenamefont {Kueng}, \citenamefont {Huang}, \citenamefont {van Bijnen}, \citenamefont {Kokail}, \citenamefont {Dalmonte}, \citenamefont {Calabrese}, \citenamefont {Kraus}, \citenamefont {Preskill}, \citenamefont {Zoller},\ and\ \citenamefont {Vermersch}}]{Elben_2020b}%
  \BibitemOpen
  \bibfield  {author} {\bibinfo {author} {\bibfnamefont {A.}~\bibnamefont {Elben}}, \bibinfo {author} {\bibfnamefont {R.}~\bibnamefont {Kueng}}, \bibinfo {author} {\bibfnamefont {H.-Y.~R.}\ \bibnamefont {Huang}}, \bibinfo {author} {\bibfnamefont {R.}~\bibnamefont {van Bijnen}}, \bibinfo {author} {\bibfnamefont {C.}~\bibnamefont {Kokail}}, \bibinfo {author} {\bibfnamefont {M.}~\bibnamefont {Dalmonte}}, \bibinfo {author} {\bibfnamefont {P.}~\bibnamefont {Calabrese}}, \bibinfo {author} {\bibfnamefont {B.}~\bibnamefont {Kraus}}, \bibinfo {author} {\bibfnamefont {J.}~\bibnamefont {Preskill}}, \bibinfo {author} {\bibfnamefont {P.}~\bibnamefont {Zoller}},\ and\ \bibinfo {author} {\bibfnamefont {B.}~\bibnamefont {Vermersch}},\ }\href {https://doi.org/10.1103/PhysRevLett.125.200501} {\bibfield  {journal} {\bibinfo  {journal} {Phys. Rev. Lett.}\ }\textbf {\bibinfo {volume} {125}},\ \bibinfo {pages} {200501} (\bibinfo {year} {2020})}\BibitemShut {NoStop}%
\bibitem [{\citenamefont {Tran}\ \emph {et~al.}(2015)\citenamefont {Tran}, \citenamefont {Daki\ifmmode~\acute{c}\else \'{c}\fi{}}, \citenamefont {Arnault}, \citenamefont {Laskowski},\ and\ \citenamefont {Paterek}}]{Tran_2015}%
  \BibitemOpen
  \bibfield  {author} {\bibinfo {author} {\bibfnamefont {M.~C.}\ \bibnamefont {Tran}}, \bibinfo {author} {\bibfnamefont {B.}~\bibnamefont {Daki\ifmmode~\acute{c}\else \'{c}\fi{}}}, \bibinfo {author} {\bibfnamefont {F.~m.~c.}\ \bibnamefont {Arnault}}, \bibinfo {author} {\bibfnamefont {W.}~\bibnamefont {Laskowski}},\ and\ \bibinfo {author} {\bibfnamefont {T.}~\bibnamefont {Paterek}},\ }\href {https://doi.org/10.1103/PhysRevA.92.050301} {\bibfield  {journal} {\bibinfo  {journal} {Phys. Rev. A}\ }\textbf {\bibinfo {volume} {92}},\ \bibinfo {pages} {050301} (\bibinfo {year} {2015})}\BibitemShut {NoStop}%
\bibitem [{\citenamefont {Tran}\ \emph {et~al.}(2016)\citenamefont {Tran}, \citenamefont {Daki\ifmmode~\acute{c}\else \'{c}\fi{}}, \citenamefont {Laskowski},\ and\ \citenamefont {Paterek}}]{Tran_2016}%
  \BibitemOpen
  \bibfield  {author} {\bibinfo {author} {\bibfnamefont {M.~C.}\ \bibnamefont {Tran}}, \bibinfo {author} {\bibfnamefont {B.}~\bibnamefont {Daki\ifmmode~\acute{c}\else \'{c}\fi{}}}, \bibinfo {author} {\bibfnamefont {W.}~\bibnamefont {Laskowski}},\ and\ \bibinfo {author} {\bibfnamefont {T.}~\bibnamefont {Paterek}},\ }\href {https://doi.org/10.1103/PhysRevA.94.042302} {\bibfield  {journal} {\bibinfo  {journal} {Phys. Rev. A}\ }\textbf {\bibinfo {volume} {94}},\ \bibinfo {pages} {042302} (\bibinfo {year} {2016})}\BibitemShut {NoStop}%
\bibitem [{\citenamefont {Dimi{\'{c}}}\ and\ \citenamefont {Daki{\'{c}}}(2018)}]{Dimi__2018}%
  \BibitemOpen
  \bibfield  {author} {\bibinfo {author} {\bibfnamefont {A.}~\bibnamefont {Dimi{\'{c}}}}\ and\ \bibinfo {author} {\bibfnamefont {B.}~\bibnamefont {Daki{\'{c}}}},\ }\href {https://doi.org/10.1038%2Fs41534-017-0055-x} {\bibfield  {journal} {\bibinfo  {journal} {npj Quantum Inf.}\ }\textbf {\bibinfo {volume} {4}},\ \bibinfo {pages} {11} (\bibinfo {year} {2018})}\BibitemShut {NoStop}%
\bibitem [{\citenamefont {Saggio}\ \emph {et~al.}(2019)\citenamefont {Saggio}, \citenamefont {Dimi{\'{c}}}, \citenamefont {Greganti}, \citenamefont {Rozema}, \citenamefont {Walther},\ and\ \citenamefont {Daki{\'{c}}}}]{Saggio_2019}%
  \BibitemOpen
  \bibfield  {author} {\bibinfo {author} {\bibfnamefont {V.}~\bibnamefont {Saggio}}, \bibinfo {author} {\bibfnamefont {A.}~\bibnamefont {Dimi{\'{c}}}}, \bibinfo {author} {\bibfnamefont {C.}~\bibnamefont {Greganti}}, \bibinfo {author} {\bibfnamefont {L.~A.}\ \bibnamefont {Rozema}}, \bibinfo {author} {\bibfnamefont {P.}~\bibnamefont {Walther}},\ and\ \bibinfo {author} {\bibfnamefont {B.}~\bibnamefont {Daki{\'{c}}}},\ }\href {https://doi.org/10.1038%2Fs41567-019-0550-4} {\bibfield  {journal} {\bibinfo  {journal} {Nat. Phys.}\ }\textbf {\bibinfo {volume} {15}},\ \bibinfo {pages} {935} (\bibinfo {year} {2019})}\BibitemShut {NoStop}%
\bibitem [{\citenamefont {Ketterer}\ \emph {et~al.}(2019)\citenamefont {Ketterer}, \citenamefont {Wyderka},\ and\ \citenamefont {G\"uhne}}]{Ketterer_2019}%
  \BibitemOpen
  \bibfield  {author} {\bibinfo {author} {\bibfnamefont {A.}~\bibnamefont {Ketterer}}, \bibinfo {author} {\bibfnamefont {N.}~\bibnamefont {Wyderka}},\ and\ \bibinfo {author} {\bibfnamefont {O.}~\bibnamefont {G\"uhne}},\ }\href {https://doi.org/10.1103/PhysRevLett.122.120505} {\bibfield  {journal} {\bibinfo  {journal} {Phys. Rev. Lett.}\ }\textbf {\bibinfo {volume} {122}},\ \bibinfo {pages} {120505} (\bibinfo {year} {2019})}\BibitemShut {NoStop}%
\bibitem [{\citenamefont {Ketterer}\ \emph {et~al.}(2020)\citenamefont {Ketterer}, \citenamefont {Wyderka},\ and\ \citenamefont {Gühne}}]{Ketterer_2020}%
  \BibitemOpen
  \bibfield  {author} {\bibinfo {author} {\bibfnamefont {A.}~\bibnamefont {Ketterer}}, \bibinfo {author} {\bibfnamefont {N.}~\bibnamefont {Wyderka}},\ and\ \bibinfo {author} {\bibfnamefont {O.}~\bibnamefont {Gühne}},\ }\href {https://doi.org/10.22331%2Fq-2020-09-16-325} {\bibfield  {journal} {\bibinfo  {journal} {Quantum}\ }\textbf {\bibinfo {volume} {4}},\ \bibinfo {pages} {325} (\bibinfo {year} {2020})}\BibitemShut {NoStop}%
\bibitem [{\citenamefont {Knips}\ \emph {et~al.}(2020)\citenamefont {Knips}, \citenamefont {Dziewior}, \citenamefont {K{\l}obus}, \citenamefont {Laskowski}, \citenamefont {Paterek}, \citenamefont {Shadbolt}, \citenamefont {Weinfurter},\ and\ \citenamefont {Meinecke}}]{Knips_2020}%
  \BibitemOpen
  \bibfield  {author} {\bibinfo {author} {\bibfnamefont {L.}~\bibnamefont {Knips}}, \bibinfo {author} {\bibfnamefont {J.}~\bibnamefont {Dziewior}}, \bibinfo {author} {\bibfnamefont {W.}~\bibnamefont {K{\l}obus}}, \bibinfo {author} {\bibfnamefont {W.}~\bibnamefont {Laskowski}}, \bibinfo {author} {\bibfnamefont {T.}~\bibnamefont {Paterek}}, \bibinfo {author} {\bibfnamefont {P.~J.}\ \bibnamefont {Shadbolt}}, \bibinfo {author} {\bibfnamefont {H.}~\bibnamefont {Weinfurter}},\ and\ \bibinfo {author} {\bibfnamefont {J.~D.~A.}\ \bibnamefont {Meinecke}},\ }\href {https://doi.org/10.1038%2Fs41534-020-0281-5} {\bibfield  {journal} {\bibinfo  {journal} {npj Quantum Inf.}\ }\textbf {\bibinfo {volume} {6}},\ \bibinfo {pages} {51} (\bibinfo {year} {2020})}\BibitemShut {NoStop}%
\bibitem [{\citenamefont {Imai}\ \emph {et~al.}(2021)\citenamefont {Imai}, \citenamefont {Wyderka}, \citenamefont {Ketterer},\ and\ \citenamefont {G\"uhne}}]{BoundImaiPRL2021}%
  \BibitemOpen
  \bibfield  {author} {\bibinfo {author} {\bibfnamefont {S.}~\bibnamefont {Imai}}, \bibinfo {author} {\bibfnamefont {N.}~\bibnamefont {Wyderka}}, \bibinfo {author} {\bibfnamefont {A.}~\bibnamefont {Ketterer}},\ and\ \bibinfo {author} {\bibfnamefont {O.}~\bibnamefont {G\"uhne}},\ }\href {https://doi.org/10.1103/PhysRevLett.126.150501} {\bibfield  {journal} {\bibinfo  {journal} {Phys. Rev. Lett.}\ }\textbf {\bibinfo {volume} {126}},\ \bibinfo {pages} {150501} (\bibinfo {year} {2021})}\BibitemShut {NoStop}%
\bibitem [{\citenamefont {Ketterer}\ \emph {et~al.}(2022)\citenamefont {Ketterer}, \citenamefont {Imai}, \citenamefont {Wyderka},\ and\ \citenamefont {Gühne}}]{Ketterer_2022}%
  \BibitemOpen
  \bibfield  {author} {\bibinfo {author} {\bibfnamefont {A.}~\bibnamefont {Ketterer}}, \bibinfo {author} {\bibfnamefont {S.}~\bibnamefont {Imai}}, \bibinfo {author} {\bibfnamefont {N.}~\bibnamefont {Wyderka}},\ and\ \bibinfo {author} {\bibfnamefont {O.}~\bibnamefont {Gühne}},\ }\href {https://doi.org/10.1103%2Fphysreva.106.l010402} {\bibfield  {journal} {\bibinfo  {journal} {Phys. Rev. A}\ }\textbf {\bibinfo {volume} {106}},\ \bibinfo {pages} {L010402} (\bibinfo {year} {2022})}\BibitemShut {NoStop}%
\bibitem [{\citenamefont {Liu}\ \emph {et~al.}(2023)\citenamefont {Liu}, \citenamefont {He}, \citenamefont {Huber}, \citenamefont {G\"uhne},\ and\ \citenamefont {Vitagliano}}]{liu2023characterizing}%
  \BibitemOpen
  \bibfield  {author} {\bibinfo {author} {\bibfnamefont {S.}~\bibnamefont {Liu}}, \bibinfo {author} {\bibfnamefont {Q.}~\bibnamefont {He}}, \bibinfo {author} {\bibfnamefont {M.}~\bibnamefont {Huber}}, \bibinfo {author} {\bibfnamefont {O.}~\bibnamefont {G\"uhne}},\ and\ \bibinfo {author} {\bibfnamefont {G.}~\bibnamefont {Vitagliano}},\ }\href {https://doi.org/10.1103/PRXQuantum.4.020324} {\bibfield  {journal} {\bibinfo  {journal} {PRX Quantum}\ }\textbf {\bibinfo {volume} {4}},\ \bibinfo {pages} {020324} (\bibinfo {year} {2023})}\BibitemShut {NoStop}%
\bibitem [{\citenamefont {Wyderka}\ and\ \citenamefont {Ketterer}(2023)}]{ProbingWyderkaPRXQ2023}%
  \BibitemOpen
  \bibfield  {author} {\bibinfo {author} {\bibfnamefont {N.}~\bibnamefont {Wyderka}}\ and\ \bibinfo {author} {\bibfnamefont {A.}~\bibnamefont {Ketterer}},\ }\href {https://doi.org/10.1103/PRXQuantum.4.020325} {\bibfield  {journal} {\bibinfo  {journal} {PRX Quantum}\ }\textbf {\bibinfo {volume} {4}},\ \bibinfo {pages} {020325} (\bibinfo {year} {2023})}\BibitemShut {NoStop}%
\bibitem [{\citenamefont {Wyderka}\ \emph {et~al.}(2023)\citenamefont {Wyderka}, \citenamefont {Ketterer}, \citenamefont {Imai}, \citenamefont {B{\"o}nsel}, \citenamefont {Jones}, \citenamefont {Kirby}, \citenamefont {Yu},\ and\ \citenamefont {G{\"u}hne}}]{wyderka2023complete}%
  \BibitemOpen
  \bibfield  {author} {\bibinfo {author} {\bibfnamefont {N.}~\bibnamefont {Wyderka}}, \bibinfo {author} {\bibfnamefont {A.}~\bibnamefont {Ketterer}}, \bibinfo {author} {\bibfnamefont {S.}~\bibnamefont {Imai}}, \bibinfo {author} {\bibfnamefont {J.~L.}\ \bibnamefont {B{\"o}nsel}}, \bibinfo {author} {\bibfnamefont {D.~E.}\ \bibnamefont {Jones}}, \bibinfo {author} {\bibfnamefont {B.~T.}\ \bibnamefont {Kirby}}, \bibinfo {author} {\bibfnamefont {X.-D.}\ \bibnamefont {Yu}},\ and\ \bibinfo {author} {\bibfnamefont {O.}~\bibnamefont {G{\"u}hne}},\ }\href {https://doi.org/10.1103/PhysRevLett.131.090201} {\bibfield  {journal} {\bibinfo  {journal} {Phys. Rev. Lett.}\ }\textbf {\bibinfo {volume} {131}},\ \bibinfo {pages} {090201} (\bibinfo {year} {2023})}\BibitemShut {NoStop}%
\bibitem [{\citenamefont {Zhang}\ \emph {et~al.}(2023)\citenamefont {Zhang}, \citenamefont {Zhao}, \citenamefont {Wyderka}, \citenamefont {Imai}, \citenamefont {Ketterer}, \citenamefont {Wang}, \citenamefont {Xu}, \citenamefont {Li}, \citenamefont {Liu}, \citenamefont {Huang} \emph {et~al.}}]{zhang2023experimental}%
  \BibitemOpen
  \bibfield  {author} {\bibinfo {author} {\bibfnamefont {C.}~\bibnamefont {Zhang}}, \bibinfo {author} {\bibfnamefont {Y.-Y.}\ \bibnamefont {Zhao}}, \bibinfo {author} {\bibfnamefont {N.}~\bibnamefont {Wyderka}}, \bibinfo {author} {\bibfnamefont {S.}~\bibnamefont {Imai}}, \bibinfo {author} {\bibfnamefont {A.}~\bibnamefont {Ketterer}}, \bibinfo {author} {\bibfnamefont {N.-N.}\ \bibnamefont {Wang}}, \bibinfo {author} {\bibfnamefont {K.}~\bibnamefont {Xu}}, \bibinfo {author} {\bibfnamefont {K.}~\bibnamefont {Li}}, \bibinfo {author} {\bibfnamefont {B.-H.}\ \bibnamefont {Liu}}, \bibinfo {author} {\bibfnamefont {Y.-F.}\ \bibnamefont {Huang}}, \emph {et~al.},\ }\href@noop {} {\bibfield  {journal} {\bibinfo  {journal} {arXiv preprint arXiv:2307.04382}\ } (\bibinfo {year} {2023})}\BibitemShut {NoStop}%
\bibitem [{sup()}]{supp}%
  \BibitemOpen
  \href@noop {} {\bibinfo {title} {See supplemental material for details.}}\BibitemShut {Stop}%
\bibitem [{\citenamefont {Lib}\ and\ \citenamefont {Bromberg}(2022)}]{lib2022quantum}%
  \BibitemOpen
  \bibfield  {author} {\bibinfo {author} {\bibfnamefont {O.}~\bibnamefont {Lib}}\ and\ \bibinfo {author} {\bibfnamefont {Y.}~\bibnamefont {Bromberg}},\ }\href {https://doi.org/https://doi.org/10.1038/s41567-022-01692-y} {\bibfield  {journal} {\bibinfo  {journal} {Nat. Phys.}\ }\textbf {\bibinfo {volume} {18}},\ \bibinfo {pages} {986} (\bibinfo {year} {2022})}\BibitemShut {NoStop}%
\bibitem [{\citenamefont {Morizur}\ \emph {et~al.}(2010)\citenamefont {Morizur}, \citenamefont {Nicholls}, \citenamefont {Jian}, \citenamefont {Armstrong}, \citenamefont {Treps}, \citenamefont {Hage}, \citenamefont {Hsu}, \citenamefont {Bowen}, \citenamefont {Janousek},\ and\ \citenamefont {Bachor}}]{morizur2010programmable}%
  \BibitemOpen
  \bibfield  {author} {\bibinfo {author} {\bibfnamefont {J.-F.}\ \bibnamefont {Morizur}}, \bibinfo {author} {\bibfnamefont {L.}~\bibnamefont {Nicholls}}, \bibinfo {author} {\bibfnamefont {P.}~\bibnamefont {Jian}}, \bibinfo {author} {\bibfnamefont {S.}~\bibnamefont {Armstrong}}, \bibinfo {author} {\bibfnamefont {N.}~\bibnamefont {Treps}}, \bibinfo {author} {\bibfnamefont {B.}~\bibnamefont {Hage}}, \bibinfo {author} {\bibfnamefont {M.}~\bibnamefont {Hsu}}, \bibinfo {author} {\bibfnamefont {W.}~\bibnamefont {Bowen}}, \bibinfo {author} {\bibfnamefont {J.}~\bibnamefont {Janousek}},\ and\ \bibinfo {author} {\bibfnamefont {H.-A.}\ \bibnamefont {Bachor}},\ }\href {https://doi.org/10.1364/JOSAA.27.002524} {\bibfield  {journal} {\bibinfo  {journal} {J. Opt. Soc. Am. A}\ }\textbf {\bibinfo {volume} {27}},\ \bibinfo {pages} {2524} (\bibinfo {year} {2010})}\BibitemShut {NoStop}%
\bibitem [{\citenamefont {Labroille}\ \emph {et~al.}(2014)\citenamefont {Labroille}, \citenamefont {Denolle}, \citenamefont {Jian}, \citenamefont {Genevaux}, \citenamefont {Treps},\ and\ \citenamefont {Morizur}}]{labroille2014efficient}%
  \BibitemOpen
  \bibfield  {author} {\bibinfo {author} {\bibfnamefont {G.}~\bibnamefont {Labroille}}, \bibinfo {author} {\bibfnamefont {B.}~\bibnamefont {Denolle}}, \bibinfo {author} {\bibfnamefont {P.}~\bibnamefont {Jian}}, \bibinfo {author} {\bibfnamefont {P.}~\bibnamefont {Genevaux}}, \bibinfo {author} {\bibfnamefont {N.}~\bibnamefont {Treps}},\ and\ \bibinfo {author} {\bibfnamefont {J.-F.}\ \bibnamefont {Morizur}},\ }\href {https://doi.org/10.1364/OE.22.015599} {\bibfield  {journal} {\bibinfo  {journal} {Opt. Express}\ }\textbf {\bibinfo {volume} {22}},\ \bibinfo {pages} {15599} (\bibinfo {year} {2014})}\BibitemShut {NoStop}%
\bibitem [{\citenamefont {Fontaine}\ \emph {et~al.}(2019)\citenamefont {Fontaine}, \citenamefont {Ryf}, \citenamefont {Chen}, \citenamefont {Neilson}, \citenamefont {Kim},\ and\ \citenamefont {Carpenter}}]{fontaine2019laguerre}%
  \BibitemOpen
  \bibfield  {author} {\bibinfo {author} {\bibfnamefont {N.~K.}\ \bibnamefont {Fontaine}}, \bibinfo {author} {\bibfnamefont {R.}~\bibnamefont {Ryf}}, \bibinfo {author} {\bibfnamefont {H.}~\bibnamefont {Chen}}, \bibinfo {author} {\bibfnamefont {D.~T.}\ \bibnamefont {Neilson}}, \bibinfo {author} {\bibfnamefont {K.}~\bibnamefont {Kim}},\ and\ \bibinfo {author} {\bibfnamefont {J.}~\bibnamefont {Carpenter}},\ }\href {https://doi.org/https://doi.org/10.1038/s41467-019-09840-4} {\bibfield  {journal} {\bibinfo  {journal} {Nat. Commun.}\ }\textbf {\bibinfo {volume} {10}},\ \bibinfo {pages} {1865} (\bibinfo {year} {2019})}\BibitemShut {NoStop}%
\bibitem [{\citenamefont {Lib}\ \emph {et~al.}(2024{\natexlab{a}})\citenamefont {Lib}, \citenamefont {Sulimany}, \citenamefont {Ben-Or},\ and\ \citenamefont {Bromberg}}]{lib2024high}%
  \BibitemOpen
  \bibfield  {author} {\bibinfo {author} {\bibfnamefont {O.}~\bibnamefont {Lib}}, \bibinfo {author} {\bibfnamefont {K.}~\bibnamefont {Sulimany}}, \bibinfo {author} {\bibfnamefont {M.}~\bibnamefont {Ben-Or}},\ and\ \bibinfo {author} {\bibfnamefont {Y.}~\bibnamefont {Bromberg}},\ }\href@noop {} {\bibfield  {journal} {\bibinfo  {journal} {arXiv preprint arXiv:2403.04210}\ } (\bibinfo {year} {2024}{\natexlab{a}})}\BibitemShut {NoStop}%
\bibitem [{\citenamefont {Brandt}\ \emph {et~al.}(2020)\citenamefont {Brandt}, \citenamefont {Hiekkam\"{a}ki}, \citenamefont {Bouchard}, \citenamefont {Huber},\ and\ \citenamefont {Fickler}}]{brandt2020high}%
  \BibitemOpen
  \bibfield  {author} {\bibinfo {author} {\bibfnamefont {F.}~\bibnamefont {Brandt}}, \bibinfo {author} {\bibfnamefont {M.}~\bibnamefont {Hiekkam\"{a}ki}}, \bibinfo {author} {\bibfnamefont {F.}~\bibnamefont {Bouchard}}, \bibinfo {author} {\bibfnamefont {M.}~\bibnamefont {Huber}},\ and\ \bibinfo {author} {\bibfnamefont {R.}~\bibnamefont {Fickler}},\ }\href {https://doi.org/10.1364/OPTICA.375875} {\bibfield  {journal} {\bibinfo  {journal} {Optica}\ }\textbf {\bibinfo {volume} {7}},\ \bibinfo {pages} {98} (\bibinfo {year} {2020})}\BibitemShut {NoStop}%
\bibitem [{\citenamefont {Hiekkam\"aki}\ and\ \citenamefont {Fickler}(2021)}]{hiekkamaki2021high}%
  \BibitemOpen
  \bibfield  {author} {\bibinfo {author} {\bibfnamefont {M.}~\bibnamefont {Hiekkam\"aki}}\ and\ \bibinfo {author} {\bibfnamefont {R.}~\bibnamefont {Fickler}},\ }\href {https://doi.org/10.1103/PhysRevLett.126.123601} {\bibfield  {journal} {\bibinfo  {journal} {Phys. Rev. Lett.}\ }\textbf {\bibinfo {volume} {126}},\ \bibinfo {pages} {123601} (\bibinfo {year} {2021})}\BibitemShut {NoStop}%
\bibitem [{\citenamefont {Lib}\ \emph {et~al.}(2022)\citenamefont {Lib}, \citenamefont {Sulimany},\ and\ \citenamefont {Bromberg}}]{lib2022processing}%
  \BibitemOpen
  \bibfield  {author} {\bibinfo {author} {\bibfnamefont {O.}~\bibnamefont {Lib}}, \bibinfo {author} {\bibfnamefont {K.}~\bibnamefont {Sulimany}},\ and\ \bibinfo {author} {\bibfnamefont {Y.}~\bibnamefont {Bromberg}},\ }\href {https://doi.org/10.1103/PhysRevApplied.18.014063} {\bibfield  {journal} {\bibinfo  {journal} {Phys. Rev. Appl.}\ }\textbf {\bibinfo {volume} {18}},\ \bibinfo {pages} {014063} (\bibinfo {year} {2022})}\BibitemShut {NoStop}%
\bibitem [{\citenamefont {Lib}\ and\ \citenamefont {Bromberg}(2024)}]{lib2024resource}%
  \BibitemOpen
  \bibfield  {author} {\bibinfo {author} {\bibfnamefont {O.}~\bibnamefont {Lib}}\ and\ \bibinfo {author} {\bibfnamefont {Y.}~\bibnamefont {Bromberg}},\ }\href@noop {} {\bibfield  {journal} {\bibinfo  {journal} {Nature Photonics}\ ,\ \bibinfo {pages} {1}} (\bibinfo {year} {2024})}\BibitemShut {NoStop}%
\bibitem [{\citenamefont {Lib}\ \emph {et~al.}(2024{\natexlab{b}})\citenamefont {Lib}, \citenamefont {Shekel},\ and\ \citenamefont {Bromberg}}]{lib2024building}%
  \BibitemOpen
  \bibfield  {author} {\bibinfo {author} {\bibfnamefont {O.}~\bibnamefont {Lib}}, \bibinfo {author} {\bibfnamefont {R.}~\bibnamefont {Shekel}},\ and\ \bibinfo {author} {\bibfnamefont {Y.}~\bibnamefont {Bromberg}},\ }\href@noop {} {\bibfield  {journal} {\bibinfo  {journal} {arXiv preprint arXiv:2409.20039}\ } (\bibinfo {year} {2024}{\natexlab{b}})}\BibitemShut {NoStop}%
\bibitem [{\citenamefont {Krenn}\ \emph {et~al.}(2015)\citenamefont {Krenn}, \citenamefont {Handsteiner}, \citenamefont {Fink}, \citenamefont {Fickler},\ and\ \citenamefont {Zeilinger}}]{krenn2015twisted}%
  \BibitemOpen
  \bibfield  {author} {\bibinfo {author} {\bibfnamefont {M.}~\bibnamefont {Krenn}}, \bibinfo {author} {\bibfnamefont {J.}~\bibnamefont {Handsteiner}}, \bibinfo {author} {\bibfnamefont {M.}~\bibnamefont {Fink}}, \bibinfo {author} {\bibfnamefont {R.}~\bibnamefont {Fickler}},\ and\ \bibinfo {author} {\bibfnamefont {A.}~\bibnamefont {Zeilinger}},\ }\href {https://doi.org/10.1073/pnas.1517574112} {\bibfield  {journal} {\bibinfo  {journal} {Proc. Natl. Acad. Sci. U.S.A.}\ }\textbf {\bibinfo {volume} {112}},\ \bibinfo {pages} {14197} (\bibinfo {year} {2015})}\BibitemShut {NoStop}%
\bibitem [{\citenamefont {Valencia}\ \emph {et~al.}(2020)\citenamefont {Valencia}, \citenamefont {Goel}, \citenamefont {McCutcheon}, \citenamefont {Defienne},\ and\ \citenamefont {Malik}}]{valencia2020unscrambling}%
  \BibitemOpen
  \bibfield  {author} {\bibinfo {author} {\bibfnamefont {N.~H.}\ \bibnamefont {Valencia}}, \bibinfo {author} {\bibfnamefont {S.}~\bibnamefont {Goel}}, \bibinfo {author} {\bibfnamefont {W.}~\bibnamefont {McCutcheon}}, \bibinfo {author} {\bibfnamefont {H.}~\bibnamefont {Defienne}},\ and\ \bibinfo {author} {\bibfnamefont {M.}~\bibnamefont {Malik}},\ }\href {https://doi.org/https://doi.org/10.1038/s41567-020-0970-1} {\bibfield  {journal} {\bibinfo  {journal} {Nat. Phys.}\ }\textbf {\bibinfo {volume} {16}},\ \bibinfo {pages} {1112} (\bibinfo {year} {2020})}\BibitemShut {NoStop}%
\bibitem [{\citenamefont {Shekel}\ \emph {et~al.}(2023)\citenamefont {Shekel}, \citenamefont {Lib}, \citenamefont {Gutiérrez-Cuevas}, \citenamefont {Popoff}, \citenamefont {Ling},\ and\ \citenamefont {Bromberg}}]{shekel2023shaping}%
  \BibitemOpen
  \bibfield  {author} {\bibinfo {author} {\bibfnamefont {R.}~\bibnamefont {Shekel}}, \bibinfo {author} {\bibfnamefont {O.}~\bibnamefont {Lib}}, \bibinfo {author} {\bibfnamefont {R.}~\bibnamefont {Gutiérrez-Cuevas}}, \bibinfo {author} {\bibfnamefont {S.~M.}\ \bibnamefont {Popoff}}, \bibinfo {author} {\bibfnamefont {A.}~\bibnamefont {Ling}},\ and\ \bibinfo {author} {\bibfnamefont {Y.}~\bibnamefont {Bromberg}},\ }\href {https://doi.org/10.1063/5.0161654} {\bibfield  {journal} {\bibinfo  {journal} {APL Photonics}\ }\textbf {\bibinfo {volume} {8}},\ \bibinfo {pages} {096109} (\bibinfo {year} {2023})}\BibitemShut {NoStop}%
\bibitem [{\citenamefont {Hu}\ \emph {et~al.}(2020)\citenamefont {Hu}, \citenamefont {Xing}, \citenamefont {Liu}, \citenamefont {He}, \citenamefont {Cao}, \citenamefont {Guo}, \citenamefont {Zhang}, \citenamefont {Zhang}, \citenamefont {Huang}, \citenamefont {Li},\ and\ \citenamefont {Guo}}]{hu2020efficient}%
  \BibitemOpen
  \bibfield  {author} {\bibinfo {author} {\bibfnamefont {X.-M.}\ \bibnamefont {Hu}}, \bibinfo {author} {\bibfnamefont {W.-B.}\ \bibnamefont {Xing}}, \bibinfo {author} {\bibfnamefont {B.-H.}\ \bibnamefont {Liu}}, \bibinfo {author} {\bibfnamefont {D.-Y.}\ \bibnamefont {He}}, \bibinfo {author} {\bibfnamefont {H.}~\bibnamefont {Cao}}, \bibinfo {author} {\bibfnamefont {Y.}~\bibnamefont {Guo}}, \bibinfo {author} {\bibfnamefont {C.}~\bibnamefont {Zhang}}, \bibinfo {author} {\bibfnamefont {H.}~\bibnamefont {Zhang}}, \bibinfo {author} {\bibfnamefont {Y.-F.}\ \bibnamefont {Huang}}, \bibinfo {author} {\bibfnamefont {C.-F.}\ \bibnamefont {Li}},\ and\ \bibinfo {author} {\bibfnamefont {G.-C.}\ \bibnamefont {Guo}},\ }\href {https://doi.org/10.1364/OPTICA.388773} {\bibfield  {journal} {\bibinfo  {journal} {Optica}\ }\textbf {\bibinfo {volume} {7}},\ \bibinfo {pages} {738} (\bibinfo {year} {2020})}\BibitemShut {NoStop}%
\bibitem [{\citenamefont {Wolterink}\ \emph {et~al.}(2016)\citenamefont {Wolterink}, \citenamefont {Uppu}, \citenamefont {Ctistis}, \citenamefont {Vos}, \citenamefont {Boller},\ and\ \citenamefont {Pinkse}}]{wolterink2016programmable}%
  \BibitemOpen
  \bibfield  {author} {\bibinfo {author} {\bibfnamefont {T.~A.~W.}\ \bibnamefont {Wolterink}}, \bibinfo {author} {\bibfnamefont {R.}~\bibnamefont {Uppu}}, \bibinfo {author} {\bibfnamefont {G.}~\bibnamefont {Ctistis}}, \bibinfo {author} {\bibfnamefont {W.~L.}\ \bibnamefont {Vos}}, \bibinfo {author} {\bibfnamefont {K.-J.}\ \bibnamefont {Boller}},\ and\ \bibinfo {author} {\bibfnamefont {P.~W.~H.}\ \bibnamefont {Pinkse}},\ }\href {https://doi.org/10.1103/PhysRevA.93.053817} {\bibfield  {journal} {\bibinfo  {journal} {Phys. Rev. A}\ }\textbf {\bibinfo {volume} {93}},\ \bibinfo {pages} {053817} (\bibinfo {year} {2016})}\BibitemShut {NoStop}%
\bibitem [{\citenamefont {Defienne}\ \emph {et~al.}(2016)\citenamefont {Defienne}, \citenamefont {Barbieri}, \citenamefont {Walmsley}, \citenamefont {Smith},\ and\ \citenamefont {Gigan}}]{defienne2016two}%
  \BibitemOpen
  \bibfield  {author} {\bibinfo {author} {\bibfnamefont {H.}~\bibnamefont {Defienne}}, \bibinfo {author} {\bibfnamefont {M.}~\bibnamefont {Barbieri}}, \bibinfo {author} {\bibfnamefont {I.~A.}\ \bibnamefont {Walmsley}}, \bibinfo {author} {\bibfnamefont {B.~J.}\ \bibnamefont {Smith}},\ and\ \bibinfo {author} {\bibfnamefont {S.}~\bibnamefont {Gigan}},\ }\href {https://doi.org/10.1126/sciadv.1501054} {\bibfield  {journal} {\bibinfo  {journal} {Sci. Adv.}\ }\textbf {\bibinfo {volume} {2}},\ \bibinfo {pages} {e1501054} (\bibinfo {year} {2016})}\BibitemShut {NoStop}%
\bibitem [{\citenamefont {Defienne}\ \emph {et~al.}(2018)\citenamefont {Defienne}, \citenamefont {Reichert},\ and\ \citenamefont {Fleischer}}]{defienne2018adaptive}%
  \BibitemOpen
  \bibfield  {author} {\bibinfo {author} {\bibfnamefont {H.}~\bibnamefont {Defienne}}, \bibinfo {author} {\bibfnamefont {M.}~\bibnamefont {Reichert}},\ and\ \bibinfo {author} {\bibfnamefont {J.~W.}\ \bibnamefont {Fleischer}},\ }\href {https://doi.org/10.1103/PhysRevLett.121.233601} {\bibfield  {journal} {\bibinfo  {journal} {Phys. Rev. Lett.}\ }\textbf {\bibinfo {volume} {121}},\ \bibinfo {pages} {233601} (\bibinfo {year} {2018})}\BibitemShut {NoStop}%
\bibitem [{\citenamefont {Lib}\ \emph {et~al.}(2020)\citenamefont {Lib}, \citenamefont {Hasson},\ and\ \citenamefont {Bromberg}}]{lib2020real}%
  \BibitemOpen
  \bibfield  {author} {\bibinfo {author} {\bibfnamefont {O.}~\bibnamefont {Lib}}, \bibinfo {author} {\bibfnamefont {G.}~\bibnamefont {Hasson}},\ and\ \bibinfo {author} {\bibfnamefont {Y.}~\bibnamefont {Bromberg}},\ }\href {https://doi.org/10.1126/sciadv.abb6298} {\bibfield  {journal} {\bibinfo  {journal} {Sci. Adv.}\ }\textbf {\bibinfo {volume} {6}},\ \bibinfo {pages} {eabb6298} (\bibinfo {year} {2020})}\BibitemShut {NoStop}%
\bibitem [{\citenamefont {Shekel}\ \emph {et~al.}(2024)\citenamefont {Shekel}, \citenamefont {Lib},\ and\ \citenamefont {Bromberg}}]{shekel2024shaping}%
  \BibitemOpen
  \bibfield  {author} {\bibinfo {author} {\bibfnamefont {R.}~\bibnamefont {Shekel}}, \bibinfo {author} {\bibfnamefont {O.}~\bibnamefont {Lib}},\ and\ \bibinfo {author} {\bibfnamefont {Y.}~\bibnamefont {Bromberg}},\ }\href {https://doi.org/10.1364/OPTICAQ.525445} {\bibfield  {journal} {\bibinfo  {journal} {Optica Quantum}\ }\textbf {\bibinfo {volume} {2}},\ \bibinfo {pages} {303} (\bibinfo {year} {2024})}\BibitemShut {NoStop}%
\bibitem [{\citenamefont {Giovannini}\ \emph {et~al.}(2013)\citenamefont {Giovannini}, \citenamefont {Romero}, \citenamefont {Leach}, \citenamefont {Dudley}, \citenamefont {Forbes},\ and\ \citenamefont {Padgett}}]{GiovanniniCharacterizationPRL2013}%
  \BibitemOpen
  \bibfield  {author} {\bibinfo {author} {\bibfnamefont {D.}~\bibnamefont {Giovannini}}, \bibinfo {author} {\bibfnamefont {J.}~\bibnamefont {Romero}}, \bibinfo {author} {\bibfnamefont {J.}~\bibnamefont {Leach}}, \bibinfo {author} {\bibfnamefont {A.}~\bibnamefont {Dudley}}, \bibinfo {author} {\bibfnamefont {A.}~\bibnamefont {Forbes}},\ and\ \bibinfo {author} {\bibfnamefont {M.~J.}\ \bibnamefont {Padgett}},\ }\href {https://doi.org/10.1103/PhysRevLett.110.143601} {\bibfield  {journal} {\bibinfo  {journal} {Phys. Rev. Lett.}\ }\textbf {\bibinfo {volume} {110}},\ \bibinfo {pages} {143601} (\bibinfo {year} {2013})}\BibitemShut {NoStop}%
\bibitem [{\citenamefont {Agnew}\ \emph {et~al.}(2011)\citenamefont {Agnew}, \citenamefont {Leach}, \citenamefont {McLaren}, \citenamefont {Roux},\ and\ \citenamefont {Boyd}}]{AgnewTomographyPRA2011}%
  \BibitemOpen
  \bibfield  {author} {\bibinfo {author} {\bibfnamefont {M.}~\bibnamefont {Agnew}}, \bibinfo {author} {\bibfnamefont {J.}~\bibnamefont {Leach}}, \bibinfo {author} {\bibfnamefont {M.}~\bibnamefont {McLaren}}, \bibinfo {author} {\bibfnamefont {F.~S.}\ \bibnamefont {Roux}},\ and\ \bibinfo {author} {\bibfnamefont {R.~W.}\ \bibnamefont {Boyd}},\ }\href {https://doi.org/10.1103/PhysRevA.84.062101} {\bibfield  {journal} {\bibinfo  {journal} {Phys. Rev. A}\ }\textbf {\bibinfo {volume} {84}},\ \bibinfo {pages} {062101} (\bibinfo {year} {2011})}\BibitemShut {NoStop}%
\bibitem [{\citenamefont {Opatrn\'y}\ \emph {et~al.}(1997)\citenamefont {Opatrn\'y}, \citenamefont {Welsch},\ and\ \citenamefont {Vogel}}]{OpatrnyLeastPRA1997}%
  \BibitemOpen
  \bibfield  {author} {\bibinfo {author} {\bibfnamefont {T.}~\bibnamefont {Opatrn\'y}}, \bibinfo {author} {\bibfnamefont {D.-G.}\ \bibnamefont {Welsch}},\ and\ \bibinfo {author} {\bibfnamefont {W.}~\bibnamefont {Vogel}},\ }\href {https://doi.org/10.1103/PhysRevA.56.1788} {\bibfield  {journal} {\bibinfo  {journal} {Phys. Rev. A}\ }\textbf {\bibinfo {volume} {56}},\ \bibinfo {pages} {1788} (\bibinfo {year} {1997})}\BibitemShut {NoStop}%
\bibitem [{\citenamefont {Banaszek}\ \emph {et~al.}(1999)\citenamefont {Banaszek}, \citenamefont {D'Ariano}, \citenamefont {Paris},\ and\ \citenamefont {Sacchi}}]{BanaszekMaximumPRA1999}%
  \BibitemOpen
  \bibfield  {author} {\bibinfo {author} {\bibfnamefont {K.}~\bibnamefont {Banaszek}}, \bibinfo {author} {\bibfnamefont {G.~M.}\ \bibnamefont {D'Ariano}}, \bibinfo {author} {\bibfnamefont {M.~G.~A.}\ \bibnamefont {Paris}},\ and\ \bibinfo {author} {\bibfnamefont {M.~F.}\ \bibnamefont {Sacchi}},\ }\href {https://doi.org/10.1103/PhysRevA.61.010304} {\bibfield  {journal} {\bibinfo  {journal} {Phys. Rev. A}\ }\textbf {\bibinfo {volume} {61}},\ \bibinfo {pages} {010304} (\bibinfo {year} {1999})}\BibitemShut {NoStop}%
\bibitem [{\citenamefont {Namiki}\ and\ \citenamefont {Tokunaga}(2012)}]{namiki2012discrete}%
  \BibitemOpen
  \bibfield  {author} {\bibinfo {author} {\bibfnamefont {R.}~\bibnamefont {Namiki}}\ and\ \bibinfo {author} {\bibfnamefont {Y.}~\bibnamefont {Tokunaga}},\ }\href {https://doi.org/10.1103/PhysRevLett.108.230503} {\bibfield  {journal} {\bibinfo  {journal} {Phys. Rev. Lett.}\ }\textbf {\bibinfo {volume} {108}},\ \bibinfo {pages} {230503} (\bibinfo {year} {2012})}\BibitemShut {NoStop}%
\end{thebibliography}%
